\documentclass[prl, twocolumn, showpacs, reprint, nofootinbib, superscriptaddress, floatfix]{revtex4-2}
\usepackage[colorlinks = True, linkcolor = blue, citecolor = blue, breaklinks = True]{hyperref}

\makeatletter
\renewcommand\@seccntformat[1]{\csname the#1\endcsname.\quad}
\makeatother

\usepackage{times}
\usepackage{xcolor}
\usepackage{graphicx}
\usepackage{subfigure}
\usepackage{setspace}
\usepackage{calc}
\usepackage{calrsfs}
\usepackage{float}
\usepackage{mathtools}
\usepackage{comment}
\usepackage{ulem}
\usepackage{amssymb,amsmath,amsfonts}
\usepackage{graphics,epsfig,subfigure}
\usepackage{color,bm,hyperref}
\usepackage{xcolor}
\usepackage{tikz}
\usepackage{yfonts}
\usepackage{verbatim}
\usetikzlibrary{shapes.geometric,arrows}
\usetikzlibrary{decorations.markings}
\DeclareMathAlphabet{\pazocal}{OMS}{zplm}{m}{n}

\definecolor{lime}{HTML}{A6CE39}
\DeclareRobustCommand{\orcidicon}{\hspace{-4pt}
\begin{tikzpicture}
\draw[lime, fill=lime] (0,0)
circle [radius=0.16]
node[white] {\hspace{0.1mm}{\fontfamily{qag}\selectfont \tiny ID}};
\draw[white, fill=white] (-0.07,0.1)
circle [radius=0.01];
\end{tikzpicture}
\hspace{-3.2mm}
}
\foreach \x in {A, ..., Z}{\expandafter\xdef\csname
orcid\x\endcsname{\noexpand\href{https://orcid.org/\csname orcidauthor\x\endcsname}
{\noexpand\orcidicon}}
}

\begin{document}

\title{Finite-size fluctuations for stochastic coupled oscillators: A general theory}
\author{Rupak Majumder\orcidA{}}
\email{rupak.majumder@tifr.res.in}
\affiliation{Department of Theoretical Physics, Tata Institute of Fundamental Research, Homi Bhabha Road, Mumbai 400005, India}
\author{Julien Barr\'e\orcidB{}}
\email{julien.barre@univ-orleans.fr}
\affiliation{Institut Denis Poisson, Universit{\'e} d'Orl{\'e}ans, Universit{\'e} de Tours and CNRS, 45067 Orl{\'e}ans, France}
\author{Shamik Gupta\orcidC{}}
\email{shamik.gupta@theory.tifr.res.in}
\affiliation{Department of Theoretical Physics, Tata Institute of Fundamental Research, Homi Bhabha Road, Mumbai 400005, India}

\begin{abstract}
Phase transitions, sharp in the thermodynamic limit, get smeared in finite systems where macroscopic order-parameter fluctuations dominate. Achieving a coherent and complete theoretical description of these fluctuations is a central challenge.
We develop a general framework to quantify these finite-size effects in synchronization transitions of generic stochastic, globally-coupled nonlinear oscillators. By applying a center-manifold reduction to the nonlinear stochastic PDE for the single-oscillator distribution in finite systems, we derive a mesoscopic description that yields the complete time evolution of the order parameter in the form of a Langevin equation. In particular, this equation provides the first closed-form steady-state distribution of the order parameter, fully capturing finite-size effects. Free from integrability constraints and the celebrated Ott–Antonsen ansatz, our theory shows excellent agreement with simulations across diverse coupling functions and frequency distributions, demonstrating broad applicability. Strikingly, it surpasses recent approaches near criticality and in the incoherent phase, where finite-size fluctuations are most pronounced.

\end{abstract}

\maketitle




Spontaneous synchronization~\cite{Pikovsky:2000} is a universal phenomenon encountered across disciplines, in physics~\cite{Wiesenfeld}, chemistry~\cite{Kuramot1976}, biology \cite{Mirollo}, in systems as diverse as neural networks~\cite{Schmidt}, power grids~\cite{Totz}, nanoelectronic platforms~\cite{Matheny}, flocking animals~\cite{Ha_2010}. Since the seminal work of Winfree and Kuramoto in the 70's \cite{Winfree,Kuramoto}, globally-coupled populations of nonlinear oscillators have served as paradigmatic models for understanding such a collective behavior. In the infinite-population limit (the thermodynamic limit, $N\to \infty$), these models often show a phase transition between a synchronized (ordered) and an incoherent (disordered) phase. Theoretical analyses of these models typically consider the thermodynamic limit, where collective dynamics is captured by deterministic mean-field equations for probability distributions over oscillator states~\cite{Strogatz:2000,ABPRS05,Gupta:2014,PR15,Gupta:2018}. This formulation facilitates analytical approaches such as linear stability, bifurcation analysis, and convenient low-dimensional reduction for the time evolution of the synchronization order parameter~\cite{OA}.


In contrast to the infinite-population limit, finite populations exhibit fluctuations in the order parameter. These fluctuations are especially pronounced near the phase transition, where they lead to rounding or smearing of the singularities associated with the transition in the infinite-$N$ limit. Finite-$N$ fluctuations are not captured by theoretical analysis  done in the infinite-$N$ limit, underscoring the need for development of analytical tools that account for finite-size effects. This is particularly relevant in light of the fact that real-world systems are inherently finite. Moreover, finite-size effects often have dramatic consequences and can fundamentally alter the system behavior—stabilizing otherwise marginally-unstable unsynchronized states~\cite{Buice:2007}, generating output fluctuations (“shot noise”) that shape collective dynamics in neural circuits~\cite{Klinshov:2022}, and producing qualitatively distinct phase transitions in adaptive networks of heterogeneous oscillators~\cite{Fialkowski:2023}. Capturing these effects poses significant analytical challenges. Here, we introduce a general and tractable framework to quantify finite-size effects, and in particular, to characterize how synchronization depends on population size near transition points.

\begin{figure}
\centering
\includegraphics[width=8cm]{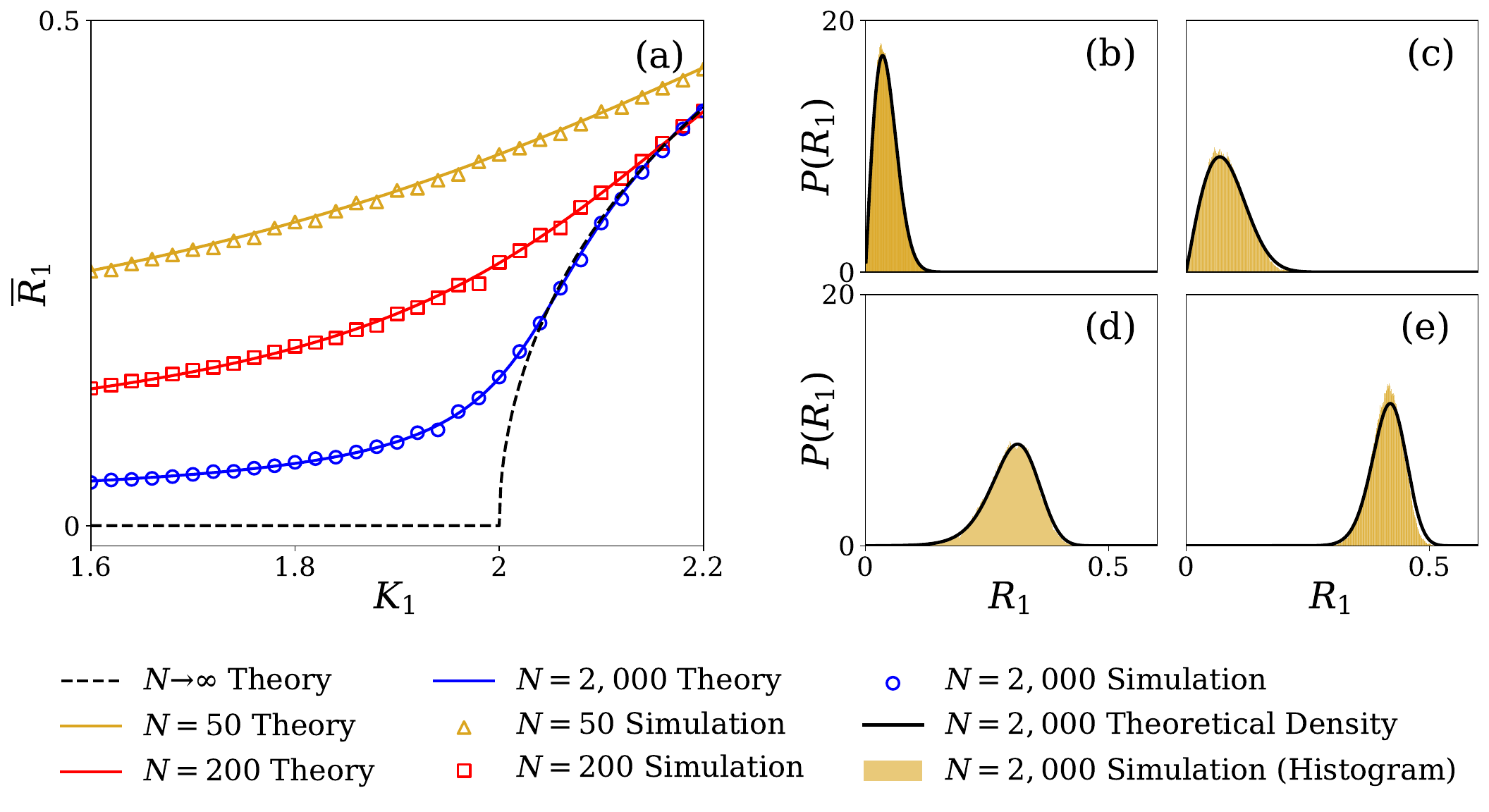} 
\caption{For model  in \textit{Application 1} with $K_2=0, D=1.0$, the figure shows agreement between our theory (lines) and numerical simulation (markers in (a) and histograms in (b) -- (e))  for the average of the steady-state order parameter in (a) and for its distribution in (b) -- (e). Parameters for (b) -- (e) are $K_1 = 1.6, 1.9, 2.1,2.2$, respectively. Simulation details for all plots are in~\cite{Simulation}. }
     \label{fig:1}
\end{figure}


In the thermodynamic limit, the collective dynamics of coupled oscillators is governed by nonlinear integro-differential PDEs for single-oscillator distribution functions. Several foundational approaches have been developed to analyze these equations in Kuramoto-like models, in particular, (i) the self-consistent equation for the order parameter introduced by Kuramoto~\cite{Kuramoto,Daido1996,Komarov2013}; (ii) bifurcation theory and center manifold reductions, which are particularly effective near the onset of synchronization~\cite{Strogatz1990,Strogatz1992,Crawford1994,Crawford1999,Chiba2015,Barre2016,Chiba2018,Gupta2018, PhysRevResearch.2.023183}; and (iii) the Ott–Antonsen (OA) ansatz~\cite{OA,Omel2018,Tyulkina18,Bick2020}, which yields exact low-dimensional reductions even far from criticality, albeit under restrictive conditions. Despite their success in the infinite-$N$ limit, extending these methods to finite populations remains a major challenge. A number of analytical and numerical efforts~\cite{Daido1990,PMT05,Chate2007,hong2015, PhysRevE.92.020901, PhysRevE.97.032310,Yue2024,Buendia2025} have been initiated to address this issue. Of particular note is the recent work of Ref.~\cite{Buendia2025} for the case of stochastic Kuramoto model, which considers the single-oscillator empirical density that in the limit $N \to \infty$ reduces to the single-oscillator distribution function. The work combines (i) the Dean–Kawasaki (DK) equation, a nonlinear stochastic PDE satisfied by the empirical density that captures finite-size fluctuations via multiplicative noise, with (ii) its reduction via the OA ansatz, to derive a single stochastic differential equation for the synchronization order parameter. While powerful, this approach relies on the applicability of the OA ansatz, originally proposed for deterministic dynamics and expected to work only approximately for stochastic dynamics with small noise. Furthermore, the applicability of the ansatz in more complex oscillator models remains to be established.


To address these issues, we propose in this Letter an alternative approach to study finite-size effects in stochastic Kuramoto-like models, by combining two major ingredients: (i) the Dean–Kawasaki equation, and (ii) an extended center manifold expansion near transition points. Our key result is Eq.~\eqref{eq:reduced}, which gives a single stochastic differential equation for the time evolution of the synchronization order parameter, while taking explicitly into account finite-$N$ effects. This equation is analytically tractable and enables direct access to both static and dynamical properties of the order parameter.


Let us view our approach in the light of  Ref.~\cite{Buendia2025} that considers the Kuramoto-Daido order parameters $Z_m = R_m e^{i\psi_m};~m \in \mathbb{Z}^+$, defined following Eq.~\eqref{eq: Generalized Kuramoto}, and uses the DK equation together with a phase ansatz $\psi_m = m\psi_1$ to derive a set of coupled evolution equations for the $R_m$'s. The evolution of $R_m$ is coupled to that of $R_{m-1}$ and $R_{m+1}$, and has a drift term $\propto 1/N$ and a stochastic correlated noise. These coupled equations are (i) used to obtain an approximate expression for average $R_1$, the synchronization order parameter, and (ii) numerically solved to obtain the the steady-state distribution of $R_1$, which are compared with simulations. A remarkable match significantly away from the transition point, even inside the synchronized phase, is reported. The agreement weakens close to the transition, where finite-$N$ fluctuations are most pronounced and which concerns us the most. From the coupled equations of $R_m$'s, the OA ansatz yields a closed equation of $R_1$, leading to greater disagreement, both at the transition and in the synchronized phase.


In contrast to Ref.~\cite{Buendia2025}, our work, which does not rely on the OA ansatz, obtains a closed evolution equation of $R_1$, Eq.~\eqref{eq:reduced}, from which we derive an explicit analytical expression for the full steady-state probability distribution of $R_1$, Eqs.~\eqref{eq: P|A|} and~\eqref{eq: PR_1}. This allows to compute the average and in fact any moment of $R_1$. Our results thus provide a complete quantitative characterization of finite-$N$ fluctuations in the order parameter, with predictions in excellent agreement with simulations in the incoherent phase and most notably near the transition point, even for $N$ as small as $50$ (see Fig.~\ref{fig:1} for the particular model studied in Ref.~\cite{Buendia2025}). Only deep in the synchronized phase, which is beyond the scope of our analysis, does disagreement appear. We extend our study to two more representative cases with increasing complexity and provide excellent agreement with simulations in all cases.


At the heart of our argument lies the observation that in the limit $N \to \infty$ and near the phase transition point, the dynamics of the single-oscillator distribution is confined to a center manifold~\cite{Crawford1994}, parameterized by the slowest decaying or unstable modes, with the fast decaying modes expressed as nonlinear functions of the former. Equation~\eqref{eq:reduced} relies on the assumption that for large enough but finite $N$, the dynamics of the empirical density remains close to the above mentioned center manifold, with finite-$N$ effects making the dynamics on this manifold stochastic. Hence, Eq.~\eqref{eq:reduced} is expected to be valid only close to the transition point and for large enough $N$, when the noise term in the equation is small. Remarkably, as mentioned above, excellent match with simulations suggests a rather broad range of validity around the transition point and even for small $N$. Though not formally rigorous in the mathematical sense, we expect our analysis to be asymptotically exact near transition points, a conjecture supported by rigorous results in related systems~\cite{ColletKraaij17,DaiPra2019, Collet2012}. Moreover, since the DK equation is related to a large deviation principle for the empirical density~\cite{DG87,DPDH96}, our method connects naturally to asymptotic analyses of large deviation principles~\cite{Barre2020,Feliachi2022}.

Turning to results, we first outline our general strategy and apply it to three increasingly complex variants of the Kuramoto model. Consider a generalized stochastic Kuramoto model of $N$ globally-coupled limit-cycle oscillators with phases $\theta_i(t) \in [0,2\pi),~i=1,2,\ldots,N$, evolving as
\begin{align}
    \frac{d \theta_i}{dt} = \omega_i + \frac{1}{N} \sum_{j = 1}^N f(\theta_j-\theta_i) + \sqrt{2D} \zeta_i(t), \label{eq: Generalized Kuramoto}
\end{align}
with Gaussian, white noise $\zeta_i(t)$ satisfying $\langle \zeta_i (t) \rangle = 0$, $\langle \zeta_i (t) \zeta_j (t') \rangle=  \delta_{ij}\delta(t-t')$, $D$ denoting noise strength, and quenched-disordered frequencies~$\omega_i$ sampled from distribution $g(\omega)$. With inter-oscillator interaction $f(\theta_j-\theta_i)~\forall~i,j$ being reciprocal and $2\pi-$periodic, we have the Fourier expansion $f(q) = \sum_{l = 1}^{\infty} K_l \sin{(lq)}$. This model exhibits phase transitions in the Kuramoto-Daido order parameters $Z_m(t) = N^{-1} \sum_{j=1}^N e^{i m \theta_j(t)};~m\in \mathbb{Z}^{+}$~\cite{PhysRevLett.77.1406,DAIDO199624}, e.g., Kuramoto model has $f(q) = K_1 \sin{q}$, with order parameter $R_1=|Z_1|$~\cite{Kuramoto}.


To describe the dynamics, one usually considers the thermodynamic limit to write a continuity equation for the single oscillator distribution $F(\theta, \omega, t)$, which takes the form of a non-linear Fokker-Planck (FP) equation. On the contrary, when $N$ is finite, the appropriate object of study is the empirical density $\bar{F}_N (\theta,\omega,t) \equiv N^{-1} \sum_{j=1}^N  \delta \left( \omega-\omega_j\right) \left( \theta-\theta_j(t)\right)$. It is known~\cite{Dean1996} that this density follows the DK equation:
\begin{align}
    \frac{\partial\bar{F}_N (\theta, \omega, t)}{\partial t} &= D\frac{\partial^2 \bar{F}_N (\theta, \omega, t) }{\partial \theta^2} - \frac{\partial }{\partial \theta} \left[\bar{F}_N (\theta, \omega, t) h\right] \nonumber\\
    &+\frac{1}{\sqrt{N}}\frac{\partial}{\partial \theta} \left[\sqrt{2D\bar{F}_N (\theta,\omega,t)}\zeta(\theta,\omega,t) \right]; \label{eq: Dean-Kawasaki}
\end{align}
$h[\bar{F}_N] \equiv \omega + \int_0^{2\pi} \int_{-\infty}^{\infty} d\theta' d\omega' f(\theta'-\theta)\bar{F}_N (\theta',\omega',t) $ and uncorrelated Gaussian, white noise field $\zeta(\theta,\omega,t)$ with $\left \langle \zeta(\theta,\omega,t)\zeta(\theta',\omega',t') \right \rangle = \delta(\theta-\theta')\delta(\omega-\omega')\delta(t-t')$. Equation~\eqref{eq: Dean-Kawasaki} is to be interpreted in the It\^{o} sense. As $N\to \infty$, when $\bar{F}_N (\theta, \omega, t) \to F(\theta, \omega, t)$, the noise term vanishes and the DK equation reduces to the FP equation for $F(\theta, \omega, t)$. We have $Z_m \equiv R_m e^{i\psi_m}=\int_0^{2\pi} \int_{-\infty}^{\infty} d\theta d\omega~e^{i m \theta }\bar{F}_N (\theta, \omega, t)$. 


We now focus on bifurcations emerging from the homogeneous stationary state, i.e., the incoherent state ($R_1=0$). In the incoherent phase, and also near the critical point within the synchronized phase, when the oscillator phases are approximately uniformly distributed over $[0,2\pi)$, we may write $\bar{F}_N (\theta, \omega, t) = \bar{g}_{_N}(\omega)/(2\pi) + \eta(\theta,\omega,t)$, where $\bar{g}_{_N}(\omega)/(2\pi)$ is the finite-$N$ incoherent state and $\eta$ denotes small fluctuations. Equation~\eqref{eq: Dean-Kawasaki} accounts for both sources of finite-$N$ fluctuations, due to (i) sampling of the $N$ frequencies from $g(\omega)$, and (ii) the stochastic noise acting on individual oscillators. Since $\bar{g}_{_N}(\omega)/(2\pi)$ approaches $g(\omega)/(2\pi)$ at large $N$, we replace $\bar{g}_N$ by $g$, thus neglecting  fluctuations in frequency sampling for further computations. The DK equation yields
 \begin{eqnarray}
    \hspace{-0.3cm}\frac{\partial \eta}{\partial t}= \mathcal{L}\eta + \mathcal{N}[\eta]+\sqrt{\frac{2D}{ N}} \frac{\partial}{\partial \theta}\left[\sqrt{\frac{g(\omega)}{2\pi}+\eta} \zeta(\theta,\omega,t)\right]. \label{eq: eta evolution}
\end{eqnarray}
Here, $\mathcal{L}$ is a linear and $\mathcal{N}$ is a nonlinear operator (Appendix A), while the third term arises from the noise field in Eq.~\eqref{eq: Dean-Kawasaki}. With $\eta(\theta,\omega,t)$ being small, $\mathcal{L}\eta$ dominates over $\mathcal{N}[\eta]$. Furthermore, if $N$ is sufficiently large such that $\mathcal{L}\eta$ also dominates over the noise in Eq.~\eqref{eq: eta evolution}, the spectrum of $\mathcal{L}$ governs the leading-order dynamics of $\eta$. In the leading order, we may consider the noise term without $\eta$, while Appendix E discusses higher-order contributions.


For any function $\Phi(\theta, \omega)$, Fourier expansion $\Phi(\theta, \omega) =\sum_{m=-\infty}^{+\infty} \Phi_m(\omega)e^{im\theta}$ leads to the convenient form 
$\mathcal{L}\Phi = \sum_{m=-\infty}^{+\infty}(L_m \Phi_m)(\omega)e^{im\theta}$. We focus on cases where an instability occurs for modes $m=\pm 1$, so that the relevant order parameter is $Z_1$. The analysis for any $m$ follows similarly. Let $\Psi_m(\theta,\omega)=\psi_m(\omega)e^{im\theta}$ be the most unstable eigenvector of $L_m$, with $\lambda_m$ the corresponding eigenvalue. As system parameters (e.g., $K_l$'s or $D$ or parameters of $g(\omega)$) are varied, $\lambda_{\pm 1}$ crossing the imaginary axis implies phase transitions.  


As $N\to \infty$ and near the transition, $\eta$ lies on the center manifold, which extends nonlinearly the subspace ${\rm Span}(\Psi_1,\Psi_{-1})$ \cite{Strogatz1990, Strogatz1992, Crawford1994, Crawford1999, Gupta2018}, and on which any function $\mathcal{F}(\theta,\omega,t)$ can be expressed as
\begin{equation}
\mathcal{F}(\theta,\omega,t) = A(t) \Psi_1 + A^\ast(t)\Psi_{-1} + W\left[A,A^*\right] \label{eq:CM ansatz},
\end{equation}
where $W\left[A,A^*\right]$ is the nonlinear contribution, parameterized by the complex number $A$, the amplitude along $\Psi_1$, and star denotes complex conjugation. Fourier expansion yields $W\left[A,A^*\right] = \sum_{l=-\infty}^{+\infty} W_l e^{il\theta}$. Exploiting the rotational symmetry of Eq.~\eqref{eq: Generalized Kuramoto} results in the following structure for the Fourier modes: $W_0 = 0,W_1 = A|A|^2 h_1\left(|A|^2\right), W_l = A^l h_l\left(|A|^2\right) $ for $l = 2,3, \ldots$, with $W_{-l} = W_l^{*}$~\cite{SM}. For large enough but finite $N$, we now assume the dynamics
of $\eta$ to remain close to the center manifold, allowing $\eta$ to be expressed in the form~\eqref{eq:CM ansatz}. Substituting Eq.~\eqref{eq:CM ansatz} into Eq.~\eqref{eq: eta evolution}, and solving order by order, we obtain a Langevin equation~\cite{SM}
\begin{align}
    \frac{dA}{dt} = \lambda_1 A-\sum_{n=1}^{\infty}c_{2n+1} A|A|^{2n} +  \sqrt{\frac{D_\mathrm{eff}}{2\pi^2 N}} \xi (t), \label{eq:reduced}
\end{align}
where we have $R_1 = 2\pi |A^{*}[1+\sum_{n=1}^{+\infty} \mathcal{A}_{1,2n}^{*}|A|^{2n}]|$ $=2\pi |A|+\mathcal{O}(|A|^3)$, and the Gaussian, white noise $\xi(t) = i \sqrt{2\pi DD_\mathrm{eff}^{-1}}\int_{-\infty}^{+\infty}d\omega\int_0^{2\pi}d\theta e^{-i\theta}\tilde{\psi}^\ast_1(\omega)\sqrt{g(\omega)}\zeta(\theta,\omega,t)$ satisfies $\left \langle \xi(t)\right \rangle = \left \langle \xi(t)\xi(t')\right \rangle= 0$, $\left \langle \xi(t) \xi^{*}(t') \right \rangle = \delta(t-t')$. The quantities $\lambda_1,c_{2n+1}, \mathcal{A}_{1,2n}, \tilde{\psi}_1(\omega)$, effective noise strength $D_\mathrm{eff}$ all depend on system parameters (Appendix~B); $c_{2n+1}$ satisfies a recursion relation, allowing computation for all $n$. Although Eq.~\eqref{eq:reduced} involves an infinite series, with $A$ small near transition, a truncated series suffices. With $A$ complex, Eq.~\eqref{eq:reduced} mimics a two-dimensional Brownian motion with a drift.

\begin{figure}
\includegraphics[width=0.7\linewidth]{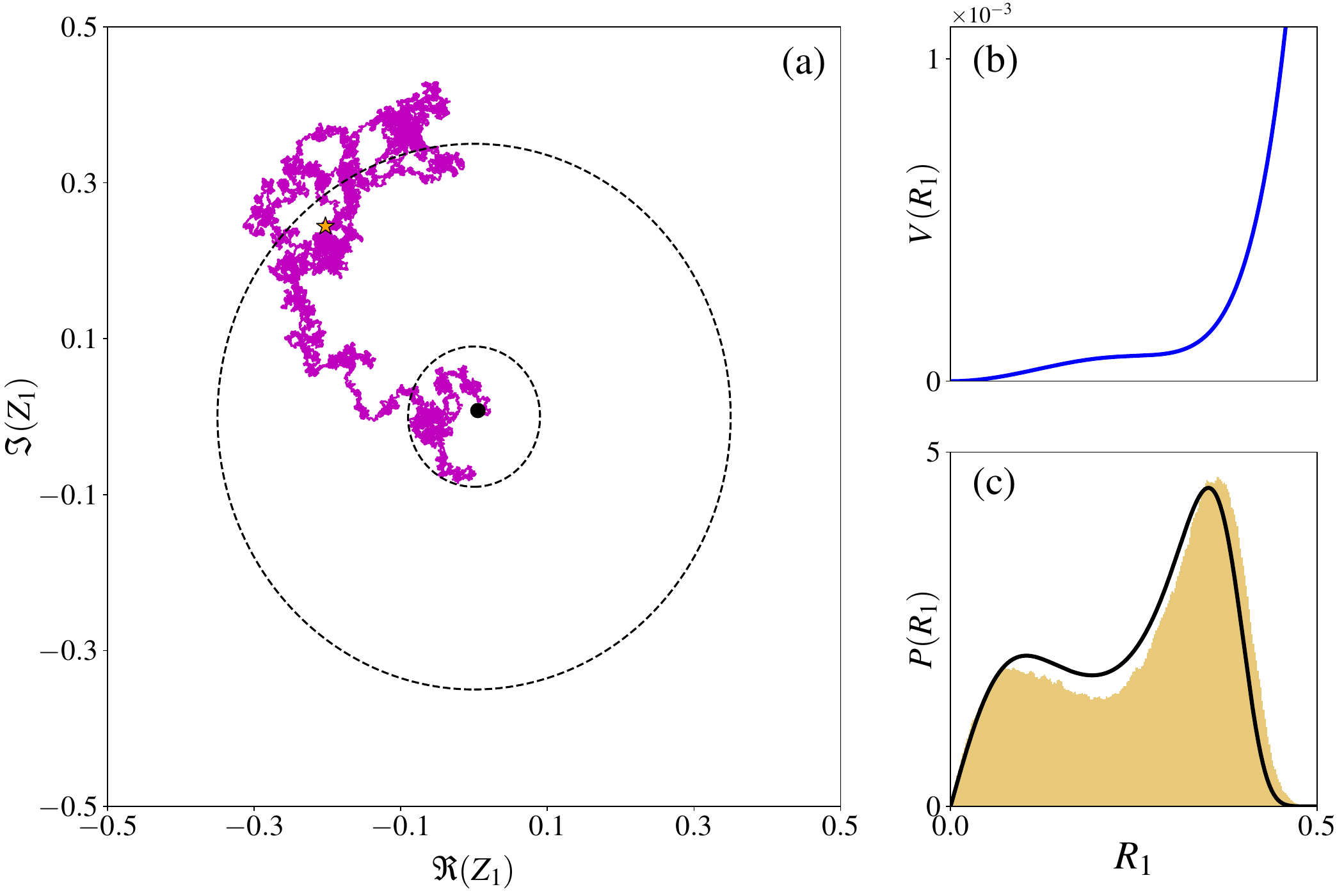}
    \caption{For model in \textit{Application 1}, the figure shows the behavior near a first-order transition point with $K_1 = 1.987, K_2 = 2.3, D=1.0, N = 10^4$. For the order parameter $R_1=|Z_1|$, panel (a) shows a numerically-obtained trajectory of $Z_1$ from $t=0.0$ (black marker) to $t=500.0$ (orange marker), displaying that it jumps between two regions that are close to the maxima of the distribution $P(R_1)$ and denoted by the two concentric circles. The effective potential $V(R_1)$ driving the dynamics of $R_1$ is shown in (b), and agreement between theory (line) and numerical simulation (histogram) is shown in (c).}
    \label{fig: 2}
\end{figure}
\begin{figure*}[htbp!]
\includegraphics[width=0.9\linewidth]{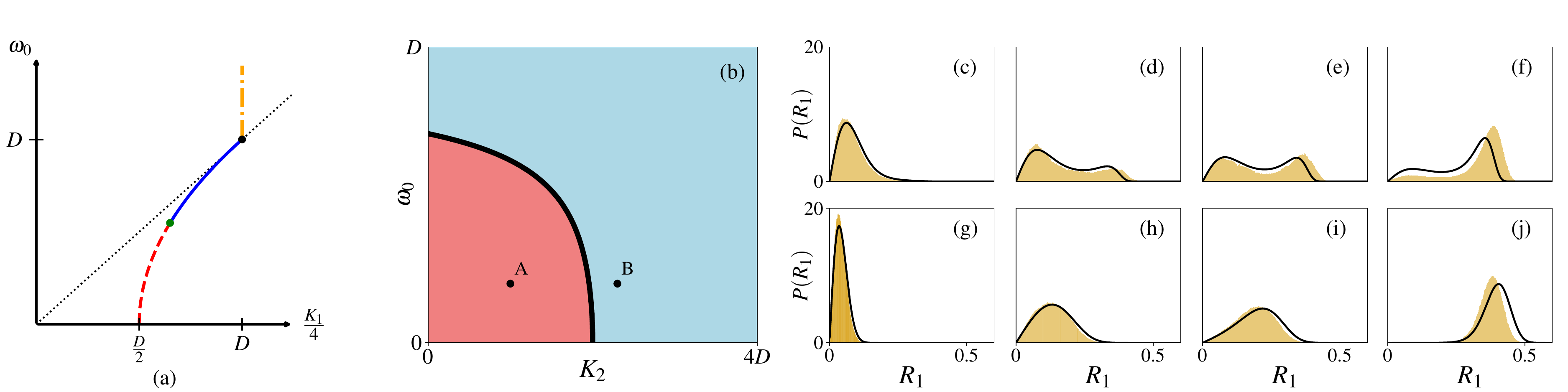}
    \caption{For model in \textit{Application 2}, (a) shows the $N\to \infty$ schematic phase diagram containing the tricritical point (green marker), continuous (red dashed line) and first-order (blue solid line) transition lines for fixed $K_2$. (b) Variation of the tricritical point (black line) with $K_2$. Upon varying $K_1$, $R_1$ shows continuous (respectively, first-order) transition for $(\omega_0,K_2)$ in red-shaded region $(\mathrm{e.g.},A\equiv(D,0.2D))$ (respectively, blue-shaded region $(\mathrm{e.g.},B\equiv(2.3D,0.2D))$). For $D=1.0$, agreement between our theory (lines) and numerical simulation (histogram) for first-order transition (point $B$ with $K_1 = 2.05, 2.059, 2.0605, 2.063$ and $N = 10^4$) is shown is (c) -- (f) and for continuous transition (point $A$ with $K_1 = 1.6, 2.059, 2.1, 2.2$ and $N = 2 \times 10^3$) is shown is (g) -- (j).}
    \label{fig:3}
\end{figure*}


Equation~\eqref{eq:reduced} is the first main result of our work: If the microscopic oscillators are driven by uncorrelated Gaussian, white noise, the order parameter near the transition evolves following Brownian dynamics with a drift. This allows to reduce the $N$-body interacting microscopic dynamics to a one-body mesoscopic description, which enables analytical predictions even for finite $N$. Notably, finite $N$ renders the order-parameter dynamics stochastic, with the noise strength scaling as $1/\sqrt{N}$. In the thermodynamic limit, when the noise vanishes, the evolution becomes deterministic, recovering all known results \cite{Strogatz1990, Strogatz1992, Crawford1994, Crawford1999, Gupta2018}. Our approach applies to both equilibrium (without frequency disorder) and non-equilibrium systems (with frequency disorder), with Eq.~\eqref{eq:reduced} taking the same universal form for both. Our approach differs from phenomenological order parameter evolution in equilibrium systems and near criticality as driven by Landau–Ginzburg free-energy functional~\cite{RevModPhys.49.435}; in contrast, we derive the time-evolution equation~\eqref{eq:reduced} from microscopic dynamics, with all parameters expressed in terms of microscopic quantities.


From Eq.~\eqref{eq:reduced}, the condition $\Re\left(\lambda_1\right) = 0$ determines the transition point as $N \to \infty$. The sign of $\Re\left(c_3\right)$ dictates the nature of the transition. For $\Re\left(c_3\right)>0$, the transition is continuous, and it suffices to consider terms up to $\mathcal{O}\left(A|A|^2\right)$. For $\Re\left(c_3\right)< 0$, the transition is first-order, requiring terms at least of  $\mathcal{O}\left(A|A|^4\right)$. In the first-order case, the system exhibits two stable states near the transition; as $N\to\infty$, the system settles into one of them depending on the initial condition. However, for finite $N$, the Brownian noise in Eq.~\eqref{eq:reduced} induces jumps between these states over time, yielding a bimodal steady-state distribution (Fig.~\ref{fig: 2} and~\ref{fig:3}(d) -- (f)).


When $\lambda_1, D_\mathrm{eff},$ and $c_{2n+1}$ are real, the deterministic part of Eq.~\eqref{eq:reduced} is the derivative of a two-dimensional potential. One then obtains a two-dimensional FP equation for the distribution of $A$, giving the steady-state distribution of $r \equiv |A|$ as
\begin{eqnarray}
   \hspace{-0.3cm} \mathbf{P}(r) &=&\mathcal{M}_0 r  e^{ -\frac{8\pi^2 N~V_0(r)}{D_\mathrm{eff}}};~V_0(r) = \sum_{n=0}^\infty \frac{c_{2n+1}r^{2(n+1)}}{2(n+1)},\label{eq: P|A|}
\end{eqnarray}
with normalization $\mathcal{M}_0$ and $c_1 \equiv -\lambda_1$~\cite{SM}. The steady-state distribution $P(R_1)$ can be obtained numerically from $\mathbf{P}(r)$ and the change of variable $R_1(r) = 2\pi r+2\pi \sum_{n=1}^{\infty}\mathcal{A}_{1,2n}r^{2n+1}$, when all $\mathcal{A}_{1,2n}$ are real. An approximate analytical form of $P(R_1)$ is obtained using $R_1 \approx 2\pi r$: $P_\approx(R_1) \sim R_1 \exp[-8\pi^2 N~V_0(R_1/2\pi)/D_\mathrm{eff}]$; it matches very well with $P(R_1)$ for regions of interest (Appendix C).


This is the second main result of our work: We provide an approximate closed-form expression of the steady-state distribution of $R_1$ and a closed-form expression of the steady-state distribution of $|A|$ for finite-$N$ systems. To compute average order parameter in the steady state, we compute the moments $\overline{r^m}~\forall~m$ of $r$ from Eq.~\eqref{eq: P|A|} and use them to compute $\bar{R}_1= \int_0^1dR_1~R_1P(R_1)= 2\pi \bar{r}+2\pi \sum_{n=1}^{\infty}\mathcal{A}_{1,2n}\overline{r^{2n+1}}$. Having provided all elements of our analysis here and in the Appendices allows to derive explicit results for specific models, which we report below (details in~\cite{SM}).


\textit{Application 1: Stochastic Kuramoto model with harmonic and bi-harmonic interaction and without frequency: } The first example we consider to showcase our theory is $g(\omega) = \delta(\omega),~f(q) = K_1 \sin{q} + K_2 \sin{2q}$. With no frequency disorder, finite-$N$ effects arise solely from stochastic noise, and the deterministic part in Eq.~\eqref{eq: Generalized Kuramoto}  writes as the derivative of a potential function, making it an equilibrium model. The operator $L_m$ has the eigenvalue $\lambda_m= |m |K_m/2 - m^2D$ for $m = \pm1,\pm 2$, and $\lambda_m = -m^2 D$ for $m = \pm3, \pm4,\ldots$. Clearly, for $K_1 < 2D$ and $K_2 < 4D$, all eigenvalues are negative, ensuring the stability of the incoherent phase in the thermodynamic limit. Increasing $K_1$ while keeping $K_2 < 4D$ causes $\lambda_{\pm1}$ to cross zero at the transition point $K_1^\mathrm{c} = 2D$, signaling a transition in $R_1$. For finite-$N$, we have Eq.~\eqref{eq:reduced} with
\begin{align}
    c_3 = 2\pi^2 K_1 \left( \frac{K_1 -  K_2 }{K_1-K_2+2D}\right),~~D_\mathrm{eff} = D, \label{eq: nofreqc3D}
\end{align}
with $c_5$ in~\cite{SM}. At the transition point, $c_3 \geq 0$ for $K_2 \leq 2D$ and $c_3<0$ for $2 D<K_2 \leq 4D$. Hence, the system undergoes a continuous transition for $K_2 \leq 2D$ and a first-order transition for $2D<K_2<4D$. For this model, $\mathcal{A}_{1,2n} =0~\forall~n~\geq1$~\cite{SM}, making $R_1  =2\pi r$ an exact relation, which gives $P_\approx(R_1) = P(R_1)$. The very good agreement between our theory and simulations for the continuous transition is shown in Fig.~\ref{fig:1} and for the first-order in Fig.~\ref{fig: 2}. Comparison of Eq.~\eqref{eq:reduced} with the results in Ref.~\cite{Buendia2025} is given in Appendix E.

\textit{Application 2: Stochastic Kuramoto model with harmonic and bi-harmonic interaction and bi-delta frequency distribution: } The second example we consider is out of equilibrium: we take $g(\omega) = [\delta(\omega-\omega_0)+\delta(\omega+\omega_0)]/2$, $f(q) = K_1 \sin{q} +K_2 \sin{2q}$. In the thermodynamic limit, the oscillator population is equally split between frequencies $+\omega_0$ and  $-\omega_0$. In this case, one can easily track the frequency-sampling fluctuations: in a particular realization, if $n$ (respectively, $N-n$) is the number of oscillators with frequency $+\omega_0$ (respectively, $-\omega_0$), the quantity $\alpha=n/N$ captures these fluctuations. For this realization, $\bar{g}_{_N}(\omega) =  \alpha \delta(\omega-\omega_0) + (1-\alpha) \delta(\omega+\omega_0)$. The operator $L_m$ has two eigenvalues $\lambda_{m,\pm} = -m^2D +
    \left(1/4\right)[K_ m\pm \sqrt{\left(K_m^2 - 16\omega_0^2\right)+i8(1-2\alpha)|m|\omega_0}];~m=\pm1,\pm2$, and $\lambda_{m,\pm} = -m^2D\pm im\omega_0$ for $m=\pm3,\pm4,\ldots$. The phase diagram of this model in the limit $N \to \infty$ is given in Fig.~\ref{fig:3}(a). Transition points of $R_1$ are obtained from the condition $\Re\left(\lambda_{1,\pm}\right)=0$, which gives $K^\mathrm{c}_1 = 2\left(D+\omega_0^2/D\right)$ for $\omega_0 \leq D$ and $K^\mathrm{c}_1=4D~\mathrm{for}~\omega_0 > D$. Note that $K^\mathrm{c}_1$ is independent of $K_2$. In the thermodynamic limit, for $\omega_0 \leq D$, the eigenvalues satisfy $\Re \left(\lambda_{\pm1,+}\right) = 0$ and $\Re\left(\lambda_{\pm1,-}\right) < 0$ at the transition point. Therefore, the unstable manifold is spanned by only $\Psi_{\pm1,+}$. Our reduced equation~\eqref{eq:reduced} thus correctly captures the finite-$N$ effects near the transition in this parameter regime (Red-dashed and blue-continuous line in Fig.~\ref{fig:3}(a)). A key result is the expression of the effective noise strength 
$D_\mathrm{eff} \propto [( K_1/4 + \sqrt{\left(K_1/4\right)^2-  \omega_0^2 } )^2 -\omega^2_0 ]^{-2}$ for $\alpha = 1/2$; its non-trivial dependence on the parameters highlights the need for a theory able to compute it precisely, which our work delivers; 
$D_\mathrm{eff}$ diverges on $K_1 = 4\omega_0$ line (black dotted line in Fig.~\ref{fig:3}(a)), reflecting the fact that on this line, all four eigenvalues $\lambda_{\pm1,\pm}$ have the same real part, and a reduced description using only two modes such as in Eq.~\eqref{eq:CM ansatz} becomes insufficient. 
Similarly, for $\omega_0>D$, all four eigenvalues satisfy $\Re\left(\lambda_{\pm1,\pm}\right) = 0$ at the transition point. As a result, the unstable manifold is spanned by all four eigenfunctions $\Psi_{\pm1,\pm}$. Such higher-dimensional structures can be tackled building on the foundation proposed in the current work.  Hence, our analysis is valid on the right of $K_1 = 4\omega_0$ line, with $0\leq\omega_0<D$.

As usual, the nature of transition in the limit $N\to \infty$ is given by the sign of $c_3$, which takes the form $c_3 \propto  \left[\left(2D^2-4\omega_0^2\right)\left(4D^2+\omega_0^2\right)-DK_2\left(4D^2-5\omega_0^2\right)\right]$. Comparison between our theory and simulations for the continuous transition is shown in Fig.~\ref{fig:3}(g) -- (j) and for first-order transition in Fig.~\ref{fig:3}(c) -- (f); the agreement is very good close to the transition point and in the incoherent phase, with small discrepancies appearing deep in the synchronized phase.

\textit{Application 3: Stochastic Kuramoto model with harmonic and bi-harmonic interaction and Lorentzian frequency distribution: } The third example we consider has a continuous frequency distribution $g(\omega) = \sigma \left[\pi \left(\omega^2+\sigma^2\right) \right]^{-1}$ and $f(q) = K_1 \sin{q} +K_2 \sin{2q}$. The operator $L_m$ has a continuous spectrum on the line $\Re(\lambda)=-Dm^2$, and for $m = \pm1,\pm 2$ and $K_m < 2\sigma$ a single eigenvalue $\lambda_m = |m |\left(K_m-2\sigma\right)/2 -m^2D$. The transition point as $N \to \infty$ is $K^\mathrm{c}_1 = 2\left(D+\sigma\right)$. For finite-$N$, we have Eq.~\eqref{eq:reduced} with 
\begin{align}
    \hspace{-0.2cm}c_3 = \frac{2\pi^2 K^2_1}{K_1-2\sigma} \left( \frac{K_1 -  K_2 -2\sigma}{K_1-K_2+2D}\right),~D_\mathrm{eff} = \frac{DK_1}{K_1-2\sigma}, \label{eq: exp lorentzian}
\end{align}
with $c_5$ in~\cite{SM}; $\sigma=0$ in Eq.~\eqref{eq: exp lorentzian} recovers the expressions in Eq.~\eqref{eq: nofreqc3D}. As noted for \textit{Application II}, the divergence of $c_3$ and $D_\mathrm{eff}$ at $K_1=2\sigma$ implies the need for a higher-dimensional reduced description, and the above analysis is valid for $K_1>2\sigma$. Comparison between our theory and simulations is presented in Fig.~\ref{fig: 4}, demonstrating a very good agreement close to the transition and in the incoherent phase.

In summary, our method provides a precise and versatile framework for dealing with finite-size effects in a wide variety of synchronization models. In our treatment, we have not considered frequency-sampling fluctuations, which are discussed in Appendix F.  Across all three applications presented, our analysis successfully captures nontrivial finite-$N$ effects in the order parameter distribution, including broad, asymmetric, and bimodal features. Its accuracy within our framework can be systematically improved by increasing the order of the expansion, or, when necessary, the dimensionality of the reduced description. This paves the way to treating more general models beyond global coupling~\cite{Medvedev2019, MedvedevTang2020, MedvedevMizuhara2021, DupuisMedvedev2022}, beyond 2D Kuramoto models~\cite{Sarthak2019, PhysRevE.104.014216}, including for instance non reciprocities \cite{fruchart2021}, as well as more complex dynamical scenarios such as metastability, where finite size effects allow rare transitions between metastable states \cite{Barre2016}. From theoretical perspectives, Eq.~\eqref{eq:reduced} can be seen as a small-noise SDE, for which there exist powerful techniques of analysis, based on Freidlin-Wentzell large deviation theory and quasi potentials \cite{Freidlin1998,Graham2005}. From experimental perspectives, potential applications lie in optical-cavity setups that inherently deal with a finite number of atoms in optical traps~\cite{PhysRevLett.113.203002, PhysRevA.92.063808, PhysRevA.95.063852, PhysRevResearch.2.013201}.

JB thanks R. Chetrite and C. Bernardin for many discussions. This research was supported by Indo-French Centre for the Promotion of Advanced Research (CEFIPRA/IFCPAR) under project identification number 6504-1. S.G. thanks ICTP–Abdus Salam International Centre for Theoretical Physics, Trieste, Italy, for
support under its Regular Associateship scheme. We gratefully acknowledge the generous allocation of computing resources by the Department of Theoretical Physics (DTP) of the Tata Institute of Fundamental Research (TIFR), and related technical assistance from Kapil Ghadiali and Ajay Salve. This work is supported by the Department of Atomic Energy, Government of India, under Project Identification Number RTI 4002. 

\bibliography{letter}

\begin{thebibliography}{10}

\bibitem{Pikovsky:2000}
Arkady Pikovsky, Michael Rosenblum, and J{\"u}rgen Kurths.
\newblock Synchronization.
\newblock {\em Cambridge university press}, 12, 2001.

\bibitem{Wiesenfeld}
Kurt Wiesenfeld, Pere Colet, and Steven~H. Strogatz.
\newblock Frequency locking in josephson arrays: Connection with the kuramoto
  model.
\newblock {\em Phys. Rev. E}, 57:1563--1569, Feb 1998.

\bibitem{Kuramot1976}
Yoshiki Kuramoto and Tomoji Yamada.
\newblock Pattern formation in oscillatory chemical reactions.
\newblock {\em Progress of Theoretical Physics}, 56:724--740, September 1976.

\bibitem{Mirollo}
Renato~E. Mirollo and Steven~H. Strogatz.
\newblock Synchronization of pulse-coupled biological oscillators.
\newblock {\em SIAM Journal on Applied Mathematics}, 50(6):1645--1662, 1990.

\bibitem{Schmidt}
R.~Schmidt, K.J.R. LaFleur, M.A. de~Reus, and et~al.
\newblock Kuramoto model simulation of neural hubs and dynamic synchrony in the
  human cerebral connectome.
\newblock {\em BMC Neuroscience}, 16(54), 2015.

\bibitem{Totz}
Carl~H. Totz, Simona Olmi, and Eckehard Sch\"oll.
\newblock Control of synchronization in two-layer power grids.
\newblock {\em Phys. Rev. E}, 102:022311, Aug 2020.

\bibitem{Matheny}
Matthew~H. Matheny, Jeffrey Emenheiser, Warren Fon, Airlie Chapman, Anastasiya
  Salova, Martin Rohden, Jarvis Li, Mathias~Hudoba de~Badyn, Márton Pósfai,
  Leonardo Duenas-Osorio, Mehran Mesbahi, James~P. Crutchfield, M.~C. Cross,
  Raissa~M. D’Souza, and Michael~L. Roukes.
\newblock Exotic states in a simple network of nanoelectromechanical
  oscillators.
\newblock {\em Science}, 363(6431):eaav7932, 2019.

\bibitem{Ha_2010}
Seung-Yeal Ha, Eunhee Jeong, and Moon-Jin Kang.
\newblock Emergent behaviour of a generalized viscek-type flocking model.
\newblock {\em Nonlinearity}, 23(12):3139, nov 2010.

\bibitem{Winfree}
Arthur~T. Winfree.
\newblock Biological rhythms and the behavior of populations of coupled
  oscillators.
\newblock {\em Journal of Theoretical Biology}, 16(1):15--42, 1967.

\bibitem{Kuramoto}
Y~Kuramoto.
\newblock {\em Chemical oscillations, waves, and turbulence}.
\newblock Springer, 1984.

\bibitem{Strogatz:2000}
Steven~H. Strogatz.
\newblock From kuramoto to crawford: exploring the onset of synchronization in
  populations of coupled oscillators.
\newblock {\em Physica D: Nonlinear Phenomena}, 143(1):1--20, 2000.

\bibitem{ABPRS05}
Juan~A. Acebr\'on, L.~L. Bonilla, Conrad~J. P\'erez~Vicente, F\'elix Ritort,
  and Renato Spigler.
\newblock The kuramoto model: A simple paradigm for synchronization phenomena.
\newblock {\em Rev. Mod. Phys.}, 77:137--185, Apr 2005.

\bibitem{Gupta:2014}
Shamik Gupta, Alessandro Campa, and Stefano Ruffo.
\newblock Kuramoto model of synchronization: equilibrium and nonequilibrium
  aspects.
\newblock {\em Journal of Statistical Mechanics: Theory and Experiment},
  2014(8):R08001, Aug 2014.

\bibitem{PR15}
Arkady Pikovsky and Michael Rosenblum.
\newblock Dynamics of globally coupled oscillators: {P}rogress and
  perspectives.
\newblock {\em Chaos: An Interdisciplinary Journal of Nonlinear Science},
  25(9), 2015.

\bibitem{Gupta:2018}
Shamik Gupta, Alessandro Campa, and Stefano Ruffo.
\newblock {\em Statistical physics of synchronization}, volume~48.
\newblock Springer, 2018.

\bibitem{OA}
Edward Ott and Thomas~M Antonsen.
\newblock Low dimensional behavior of large systems of globally coupled
  oscillators.
\newblock {\em Chaos: An Interdisciplinary Journal of Nonlinear Science},
  18(3):037113, September 2008.

\bibitem{Buice:2007}
Michael~A. Buice and Carson~C. Chow.
\newblock Correlations, fluctuations, and stability of a finite-size network of
  coupled oscillators.
\newblock {\em Phys. Rev. E}, 76:031118, Sep 2007.

\bibitem{Klinshov:2022}
Vladimir~V. Klinshov and Sergey~Yu. Kirillov.
\newblock Shot noise in next-generation neural mass models for finite-size
  networks.
\newblock {\em Phys. Rev. E}, 106:L062302, Dec 2022.

\bibitem{Fialkowski:2023}
Jan Fialkowski, Serhiy Yanchuk, Igor~M. Sokolov, Eckehard Sch\"oll, Georg~A.
  Gottwald, and Rico Berner.
\newblock Heterogeneous nucleation in finite-size adaptive dynamical networks.
\newblock {\em Phys. Rev. Lett.}, 130:067402, Feb 2023.

\bibitem{Simulation}
In the simulation the determinatic part is evolved using fourth order
  {R}ange-{K}utta method and the stochastic noise is simulated using
  {E}uler–{M}aruyama method. {F}or ${N} = 10000$, the time step is chosen
  0.01 and for ${N} = 2000$, the time step is chosen to be $0.1$.

\bibitem{Daido1996}
Hiroaki Daido.
\newblock Multibranch entrainment and scaling in large populations of coupled
  oscillators.
\newblock {\em Phys. Rev. Lett.}, 77:1406--1409, Aug 1996.

\bibitem{Komarov2013}
Maxim Komarov and Arkady Pikovsky.
\newblock Multiplicity of singular synchronous states in the kuramoto model of
  coupled oscillators.
\newblock {\em Phys. Rev. Lett.}, 111:204101, Nov 2013.

\bibitem{Strogatz1990}
Steven~H Strogatz and Renato~E Mirollo.
\newblock Stability of incoherence in a population of coupled oscillators.
\newblock {\em Journal of Statistical Physics}, 63:613--635, 1991.

\bibitem{Strogatz1992}
Steven~H. Strogatz, Renato~E. Mirollo, and Paul~C. Matthews.
\newblock Coupled nonlinear oscillators below the synchronization threshold:
  Relaxation by generalized landau damping.
\newblock {\em Phys. Rev. Lett.}, 68:2730--2733, May 1992.

\bibitem{Crawford1994}
John~David Crawford.
\newblock Amplitude expansions for instabilities in populations of
  globally-coupled oscillators.
\newblock {\em Journal of statistical physics}, 74:1047--1084, 1994.

\bibitem{Crawford1999}
John~D. Crawford and K.T.R. Davies.
\newblock Synchronization of globally coupled phase oscillators: singularities
  and scaling for general couplings.
\newblock {\em Physica D: Nonlinear Phenomena}, 125(1):1--46, 1999.

\bibitem{Chiba2015}
Hayato Chiba.
\newblock A proof of the kuramoto conjecture for a bifurcation structure of the
  infinite-dimensional kuramoto model.
\newblock {\em Ergodic Theory and Dynamical Systems}, 35(3):762–834, 2015.

\bibitem{Barre2016}
J.~Barr\'e and D.~M\'etivier.
\newblock Bifurcations and singularities for coupled oscillators with inertia
  and frustration.
\newblock {\em Phys. Rev. Lett.}, 117:214102, November 2016.

\bibitem{Chiba2018}
Hayato Chiba, Georgi~S Medvedev, and Matthew~S Mizuhara.
\newblock Bifurcations in the kuramoto model on graphs.
\newblock {\em Chaos: an interdisciplinary journal of nonlinear science},
  28(7), 2018.

\bibitem{Gupta2018}
David M{\'e}tivier and Shamik Gupta.
\newblock Bifurcations in the time-delayed kuramoto model of coupled
  oscillators: Exact results.
\newblock {\em Journal of Statistical Physics}, 176(2):279--298, 2019.

\bibitem{PhysRevResearch.2.023183}
David M\'etivier, Lucas Wetzel, and Shamik Gupta.
\newblock Onset of synchronization in networks of second-order kuramoto
  oscillators with delayed coupling: Exact results and application to
  phase-locked loops.
\newblock {\em Phys. Rev. Res.}, 2:023183, May 2020.

\bibitem{Omel2018}
Oleh~E Omel’chenko.
\newblock The mathematics behind chimera states.
\newblock {\em Nonlinearity}, 31(5):R121, 2018.

\bibitem{Tyulkina18}
Irina~V. Tyulkina, Denis~S. Goldobin, Lyudmila~S. Klimenko, and Arkady
  Pikovsky.
\newblock Dynamics of noisy oscillator populations beyond the ott-antonsen
  ansatz.
\newblock {\em Phys. Rev. Lett.}, 120:264101, Jun 2018.

\bibitem{Bick2020}
Christian Bick, Marc Goodfellow, Carlo~R Laing, and Erik~A Martens.
\newblock Understanding the dynamics of biological and neural oscillator
  networks through exact mean-field reductions: a review.
\newblock {\em The Journal of Mathematical Neuroscience}, 10(1):9, 2020.

\bibitem{Daido1990}
Hiroaki Daido.
\newblock Intrinsic fluctuations and a phase transition in a class of large
  populations of interacting oscillators.
\newblock {\em Journal of Statistical Physics}, 60:753--800, 1990.

\bibitem{PMT05}
Oleksandr~V. Popovych, Yuri~L. Maistrenko, and Peter~A. Tass.
\newblock Phase chaos in coupled oscillators.
\newblock {\em Phys. Rev. E}, 71:065201, Jun 2005.

\bibitem{Chate2007}
Hyunsuk Hong, Hugues Chat\'e, Hyunggyu Park, and Lei-Han Tang.
\newblock Entrainment transition in populations of random frequency
  oscillators.
\newblock {\em Phys. Rev. Lett.}, 99:184101, Oct 2007.

\bibitem{hong2015}
Hyunsuk Hong, Hugues Chat{\'e}, Lei-Han Tang, and Hyunggyu Park.
\newblock Finite-size scaling, dynamic fluctuations, and hyperscaling relation
  in the kuramoto model.
\newblock {\em Physical Review E}, 92(2):022122, 2015.

\bibitem{PhysRevE.92.020901}
Maxim Komarov and Arkady Pikovsky.
\newblock Finite-size-induced transitions to synchrony in oscillator ensembles
  with nonlinear global coupling.
\newblock {\em Phys. Rev. E}, 92:020901, Aug 2015.

\bibitem{PhysRevE.97.032310}
Franziska Peter and Arkady Pikovsky.
\newblock Transition to collective oscillations in finite kuramoto ensembles.
\newblock {\em Phys. Rev. E}, 97:032310, Mar 2018.

\bibitem{Yue2024}
Wenqi Yue and Georg~A. Gottwald.
\newblock A stochastic approximation for the finite-size kuramoto–sakaguchi
  model.
\newblock {\em Physica D: Nonlinear Phenomena}, 468:134292, 2024.

\bibitem{Buendia2025}
Victor Buend\'{\i}a.
\newblock Mesoscopic theory for coupled stochastic oscillators.
\newblock {\em Phys. Rev. Lett.}, 134:197201, May 2025.

\bibitem{ColletKraaij17}
Francesca Collet and Richard~C. Kraaij.
\newblock Dynamical moderate deviations for the curie–weiss model.
\newblock {\em Stochastic Processes and their Applications}, 127(9):2900--2925,
  2017.

\bibitem{DaiPra2019}
Paolo Dai~Pra and Daniele Tovazzi.
\newblock The dynamics of critical fluctuations in asymmetric curie--weiss
  models.
\newblock {\em Stochastic Processes and their Applications}, 129(3):1060--1095,
  2019.

\bibitem{Collet2012}
Francesca Collet and Paolo Dai~Pra.
\newblock The role of disorder in the dynamics of critical fluctuations of mean
  field models.
\newblock {\em Electronic Journal of Probability}, 17(26):1--40, 2012.

\bibitem{DG87}
Donald~A Dawson and J{\"u}rgen G{\"a}rtner.
\newblock Large deviations from the mckean-vlasov limit for weakly interacting
  diffusions.
\newblock {\em Stochastics: An International Journal of Probability and
  Stochastic Processes}, 20(4):247--308, 1987.

\bibitem{DPDH96}
Paolo~Dai Pra and Frank~den Hollander.
\newblock Mckean-vlasov limit for interacting random processes in random media.
\newblock {\em Journal of statistical physics}, 84:735--772, 1996.

\bibitem{Barre2020}
Julien Barr{\'e}, Cedric Bernardin, Rapha{\"e}l Ch{\'e}trite, Yash Chopra, and
  Mauro Mariani.
\newblock From fluctuating kinetics to fluctuating hydrodynamics: a
  $\gamma$-convergence of large deviations functionals approach.
\newblock {\em Journal of Statistical Physics}, 180(1):1095--1127, 2020.

\bibitem{Feliachi2022}
Ouassim Feliachi, Marc Besse, Cesare Nardini, and Julien Barr{\'e}.
\newblock Fluctuating kinetic theory and fluctuating hydrodynamics of aligning
  active particles: the dilute limit.
\newblock {\em Journal of Statistical Mechanics: Theory and Experiment},
  2022(11):113207, 2022.

\bibitem{PhysRevLett.77.1406}
Hiroaki Daido.
\newblock Multibranch entrainment and scaling in large populations of coupled
  oscillators.
\newblock {\em Phys. Rev. Lett.}, 77:1406--1409, Aug 1996.

\bibitem{DAIDO199624}
Hiroaki Daido.
\newblock Onset of cooperative entrainment in limit-cycle oscillators with
  uniform all-to-all interactions: bifurcation of the order function.
\newblock {\em Physica D: Nonlinear Phenomena}, 91(1):24--66, 1996.

\bibitem{Dean1996}
David~S Dean.
\newblock Langevin equation for the density of a system of interacting langevin
  processes.
\newblock {\em Journal of Physics A: Mathematical and General}, 29(24):L613,
  1996.

\bibitem{SM}
See supplemental material for the derivation of {E}qs.~(5) and~(6) of the main
  text, all the results for the model in applications 1, 2, and 3, and the
  derivation of {E}q.~(13) of the main text.

\bibitem{RevModPhys.49.435}
P.~C. Hohenberg and B.~I. Halperin.
\newblock Theory of dynamic critical phenomena.
\newblock {\em Rev. Mod. Phys.}, 49:435--479, Jul 1977.

\bibitem{Medvedev2019}
Georgi~S. Medvedev.
\newblock The continuum limit of the kuramoto model on sparse random graphs.
\newblock {\em Communications in Mathematical Sciences}, 17(4):883--898, 2019.

\bibitem{MedvedevTang2020}
Georgi~S. Medvedev and Xiaoya Tang.
\newblock The kuramoto model on power law graphs: Synchronization and contrast
  states.
\newblock {\em Journal of Nonlinear Science}, 30:2405--2427, 2020.

\bibitem{MedvedevMizuhara2021}
Georgi~S. Medvedev and Michael~S. Mizuhara.
\newblock Stability of clusters in the second-order kuramoto model on random
  graphs.
\newblock {\em Journal of Statistical Physics}, 182(30), 2021.

\bibitem{DupuisMedvedev2022}
Paul Dupuis and Georgi~S. Medvedev.
\newblock The large deviation principle for interacting dynamical systems on
  random graphs.
\newblock {\em Communications in Mathematical Physics}, 390:545--575, 2022.

\bibitem{Sarthak2019}
Sarthak Chandra, Michelle Girvan, and Edward Ott.
\newblock Continuous versus discontinuous transitions in the $d$-dimensional
  generalized kuramoto model: Odd $d$ is different.
\newblock {\em Phys. Rev. X}, 9:011002, Jan 2019.

\bibitem{PhysRevE.104.014216}
Chunming Zheng, Ralf Toenjes, and Arkady Pikovsky.
\newblock Transition to synchrony in a three-dimensional swarming model with
  helical trajectories.
\newblock {\em Phys. Rev. E}, 104:014216, Jul 2021.

\bibitem{fruchart2021}
Michel Fruchart, Ryo Hanai, Peter~B Littlewood, and Vincenzo Vitelli.
\newblock Non-reciprocal phase transitions.
\newblock {\em Nature}, 592(7854):363--369, 2021.

\bibitem{Freidlin1998}
Mark~Iosifovich Freidlin, Alexander~D Wentzell, MI~Freidlin, and AD~Wentzell.
\newblock {\em Random perturbations}.
\newblock Springer, 1998.

\bibitem{Graham2005}
Robert Graham.
\newblock Macroscopic potentials, bifurcations and noise in dissipative
  systems.
\newblock In {\em Fluctuations and Stochastic Phenomena in Condensed Matter:
  Proceedings of the Sitges Conference on Statistical Mechanics Sitges,
  Barcelona/Spain, May 26--30, 1986}, pages 1--34. Springer, 2005.

\bibitem{PhysRevLett.113.203002}
Stefan Sch\"utz and Giovanna Morigi.
\newblock Prethermalization of atoms due to photon-mediated long-range
  interactions.
\newblock {\em Phys. Rev. Lett.}, 113:203002, Nov 2014.

\bibitem{PhysRevA.92.063808}
Stefan Sch\"utz, Simon~B. J\"ager, and Giovanna Morigi.
\newblock Thermodynamics and dynamics of atomic self-organization in an optical
  cavity.
\newblock {\em Phys. Rev. A}, 92:063808, Dec 2015.

\bibitem{PhysRevA.95.063852}
Simon~B. J\"ager, Minghui Xu, Stefan Sch\"utz, M.~J. Holland, and Giovanna
  Morigi.
\newblock Semiclassical theory of synchronization-assisted cooling.
\newblock {\em Phys. Rev. A}, 95:063852, Jun 2017.

\bibitem{PhysRevResearch.2.013201}
Karl Pelka, Vittorio Peano, and Andr\'e Xuereb.
\newblock Chimera states in small optomechanical arrays.
\newblock {\em Phys. Rev. Res.}, 2:013201, Feb 2020.

\bibitem{carr1981centre}
Jack Carr.
\newblock {\em Applications of Centre Manifold Theory}, volume~35 of {\em
  Applied Mathematical Sciences}.
\newblock Springer, New York, NY, 1 edition, 1981.
\newblock Springer Book Archive. Springer-Verlag New York Inc. 1982.

\bibitem{Martin2021}
David Martin, Hugues Chat{\'e}, Cesare Nardini, Alexandre Solon, Julien
  Tailleur, and Fr{\'e}d{\'e}ric Van~Wijland.
\newblock Fluctuation-induced phase separation in metric and topological models
  of collective motion.
\newblock {\em Physical Review Letters}, 126(14):148001, 2021.

\end{thebibliography}
\bibliographystyle{unsrt}

\clearpage
\appendix
\setcounter{equation}{0}

\begin{figure}[htbp!]
\includegraphics[width=0.65\linewidth]{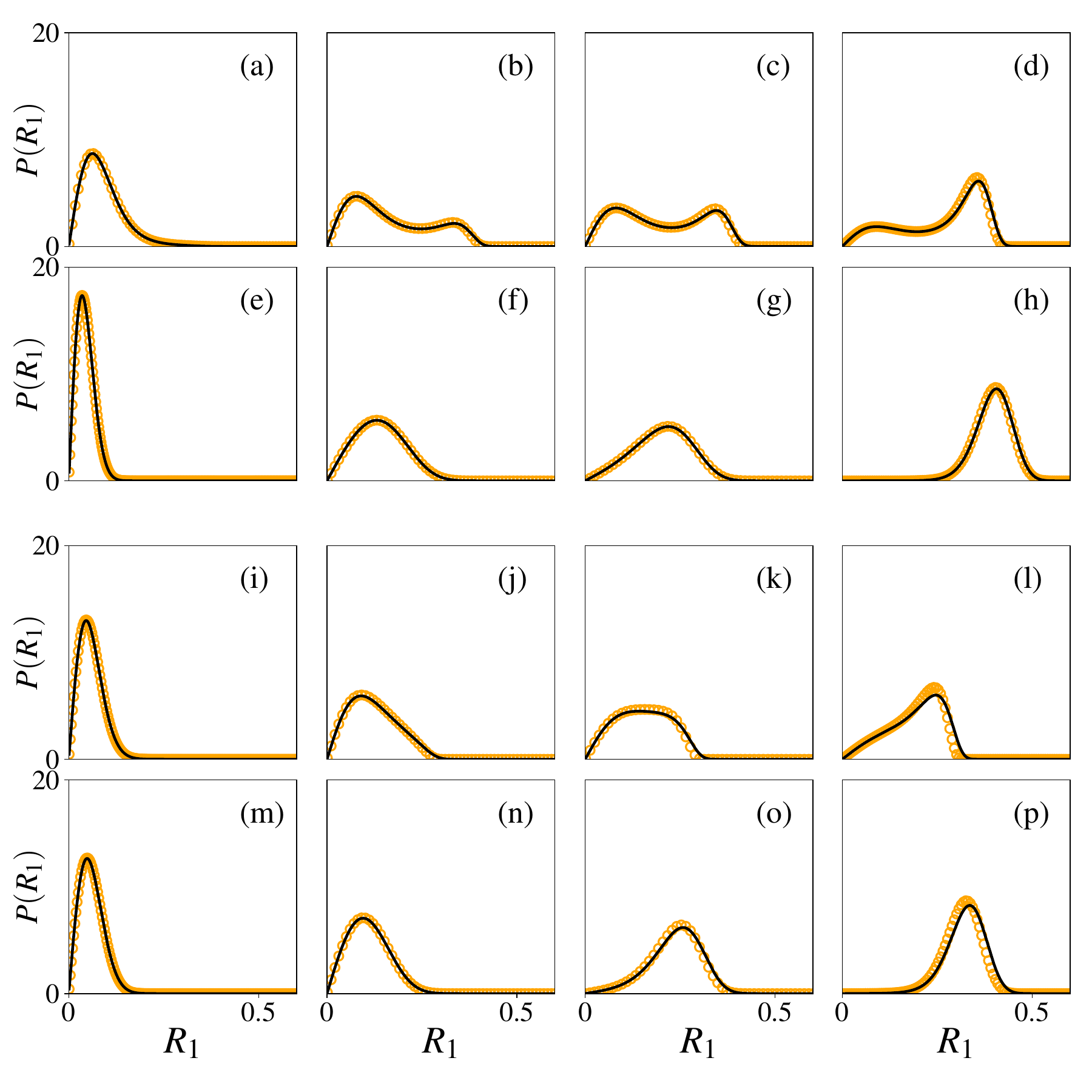}
    \caption{Agreement between $P_\approx(R_1)$ (line) and $P(R_1)$ (unfilled markers) for the model  in \textit{Application 2} ((a) -- (h)) and the one in \textit{Application 3} ((i) -- (p)) is shown. Parameters for (a) -- (h) are the same as in Fig.~\ref{fig:3}, panels (c) -- (j), respectively. Similarly, parameters for (i) -- (p) are the same as in Fig.~\ref{fig: 4}, panels (c) -- (j), respectively.}
    \label{fig: 5}
\end{figure}

\begin{center}
    \textbf{End Matter}
\end{center}
\vspace{-0.3cm}
\setcounter{equation}{0}
\renewcommand{\theequation}{EM\arabic{equation}}
\begin{center}
    \textbf{\small Appendix A: Expressions of $\mathcal{L}\eta$~and~$\mathcal{N}[\eta]$ --}
\end{center}
\vspace{-0.2cm}
The terms $\mathcal{L}\eta$ and $\mathcal{N}[\eta]$ of Eq.~\eqref{eq: eta evolution} read as
\begin{eqnarray}
    \mathcal{L}\eta &=&D\frac{\partial^2 \eta}{\partial \theta^2} -\omega\frac{\partial \eta }{\partial \theta}  - \frac{g(\omega)}{2\pi} \nonumber \\
     &&\times\int_0^{2\pi} d\theta' \int_{-\infty}^{\infty} d\omega' \partial_\theta f(\theta'-\theta)  \eta(\theta',\omega',t), \\
    \mathcal{N}[\eta] &=& -\frac{\partial}{\partial \theta}\Bigg[\eta(\theta,\omega,t) \nonumber \\
    && \times\int_0^{2\pi} d\theta' \int_{-\infty}^{\infty} d\omega' f(\theta'-\theta)\eta(\theta',\omega',t) \Bigg]. 
\end{eqnarray}
\begin{center}
    \textbf{\small Appendix B: Expressions of $\lambda_1,~c_{2n+1},~D_\mathrm{eff}$ (for derivation, see~\cite{SM}) --}
\end{center}
Let $\Psi_1(\theta,\omega) = \psi_1(\omega)e^{i\theta}$ be the eigenfunction of $\mathcal{L}$ with eigenvalue $\lambda_1$, given by the root of the function $\Lambda_1(x)$ defined as $\Lambda_1(x) \equiv 1- (K_{1}/2)\int_{-\infty}^{\infty} d\omega  g(\omega)/\left(x+D+i\omega\right)$, where we have $\psi_1(\omega) = K_{1} g(\omega)/[2\left(\lambda_1+D+i\omega\right)]$. Similarly, $\tilde{\Psi}_1(\theta,\omega) = (2\pi)^{-1}\tilde{\psi}_1(\omega)e^{i\theta}$ is the eigenfunction of $\mathcal{L}^\dagger$ with eigenvalue $\lambda^{*}_1$, with $\tilde{\psi}_{1}(\omega) = \left[[\Lambda'_1 (\lambda_{1})\right]^{*} (\lambda_1^{*}+D-i\omega)]^{-1}$. The effective noise strength is given by
\begin{equation}
    D_\mathrm{eff} = D\int_{-\infty}^{+\infty} d\omega\left|\tilde{\psi}_{1}(\omega')\right|^2g(\omega).
\end{equation}
The expression for the coefficients $c_{2n+1}$ is given by
\begin{eqnarray}
   \nonumber c_{2n+1} &=& \pi \sum_{l=1}^{\infty} K_l\sum_{q=0}^{n-l+1}\left[\mathcal{B}_{l+1,2q}\mathcal{A}^{*}_{l,2(n-l-q)} \Theta(n-l-q)\right.\\
    &&~~~~~~~~~~~~~~~~~~~~~~~~~ \left.-\mathcal{C}^{*}_{l-1,2q}\mathcal{A}_{l,2(n-l+1-q)}\right],
\end{eqnarray}
for $n=1,2,\ldots$ and $\Theta(x) = 1~\forall~x\geq0$ and $\Theta(x)=0~\forall~x<0$. Here, we have $\mathcal{A}_{m,2p} \equiv \int_{-\infty}^{\infty} d\omega ~w_{m,2p},~\mathcal{B}_{m,2p} \equiv \int_{-\infty}^{\infty} d\omega ~\tilde{\psi}^{*}_1~w_{m,2p},~\mathcal{C}_{m,2p} \equiv \int_{-\infty}^{\infty} d\omega ~\tilde{\psi}_1~w_{m,2p}$ with $c_1 \equiv -\lambda_1$, $w_{0,2p} = 0~\forall~p$, $w_{1,0} \equiv \psi_1$, while for $m>0$, we have
\begin{eqnarray}
    \nonumber w_{m,2p}=&&\left[ \frac{1}{(m+p)\lambda_1+p\lambda_1^{*}+m^2D+im\omega}\right]  \\
  \nonumber  &&\times \left[ \sum_{q=0}^{p-1}\left[(m+q)c_{2(p-q)+1}+qc^{*}_{2(p-q)+1}\right]w_{m,2q} \right.\\
  \nonumber && +m\pi \sum_{l=1}^{m}K_l\sum_{q=0}^{p} w_{m-l,2q} \mathcal{A}_{l,2(p-q)}\\
  \nonumber &&+m\pi \sum_{l=m+1}^{\infty}K_l\sum_{q=0}^{m+p-l} w^{*}_{l-m,2q} \mathcal{A}_{l,2(m+p-l-q)}\\
  \nonumber&&-m\pi \sum_{l=1}^{\infty}K_l\sum_{q=0}^{p-l} w_{m+l,2q} \mathcal{A}_{l,2(p-l-q)} \\
   && \left. + \frac{m K_m}{2} g(\omega) \mathcal{A}_{m,2p}\right].
\end{eqnarray}
\begin{center}
    \textbf{\small Appendix C: Comparison between $P(R_1)$ and $P_\approx(R_1)$ --}
\end{center}
From the expansion $R_1(r) = 2\pi r+2\pi \sum_{n=1}^{\infty}\mathcal{A}_{1,2n}r^{2n+1}$, approximating $R_1$ as $R_1 \approx 2\pi r$, we obtain an approximate analytical expression of $P(R_1)=\mathbf{P}(r)/(dR_1/dr)$ may be obtained as $P_\approx(R_1)=\mathbf{P}(R_1/(2\pi))/(2\pi)$, which on using Eq.~\eqref{eq: P|A|} reads as
\begin{equation}
    P_\approx(R_1) =\mathcal{M} R_1 e^{ -\frac{2N V(R_1)}{D_\mathrm{eff}}};~V(R_1) = \sum_{n=0}^\infty \frac{c_{2n+1}R_1^{2(n+1)}}{2(n+1)(2\pi)^{2n}}.\label{eq: PR_1}
\end{equation}
In general, $P_\approx(R_1)$ could deviate from $P(R_1)$. Remarkably, near the phase transition, which is also our region of interest, $P_\approx(R_1)$ and $P(R_1)$ agree quite well (Fig.~\ref{fig: 5}).
\begin{center}
    \textbf{\small Appendix D: Application 1, effective SDE for $R_1$}
\end{center}
The model studied in~Ref.~\cite{Buendia2025} is the model considered in \textit{Application 1} of our Letter for the particular case $K_2=0$. Putting $K_2=0$ in Eq.~\eqref{eq: nofreqc3D}, we obtain from Eq.~\eqref{eq:reduced} by retaining terms up to $\mathcal{O}(A|A|^2)$ that
\begin{align}
    \frac{dA}{dt} = \left[\frac{K_2}{2}-D\right] A -\left[\frac{2\pi^2K^2_1}{K_1+2D}\right] A|A|^{2} +  \sqrt{\frac{D}{2\pi^2 N}} \xi (t). \label{eq:reduced model 1}
\end{align}
As discussed in the main text, for this model, we have $R_1 = 2\pi |A|$. Noting that $R_1 = 2\pi \sqrt{AA^{*}}$, and using Taylor expansion of $R_1$ up to second order and the It\^{o} calculus, we obtain the evolution of $R_1$ from Eq.~\eqref{eq:reduced model 1} as
\begin{equation}
  \frac{dR_1}{dt} = \left[\frac{K_1}{2}-D\right]R_1-\frac{1}{2}\left[\frac{K^2_1}{K_1+2D}\right]R_1^3+\frac{D}{NR_1}+\sqrt{\frac{D}{N}}\xi_r(t),
\label{eqr1}
\end{equation}
with $\langle\xi_r(t)\rangle=0$ and $\langle \xi_r(t)\xi_r(t')\rangle = \delta(t-t')$. If we had retained the $\eta$ term in the noise appearing in Eq.~\eqref{eq: eta evolution} (see the main text following Eq.~\eqref{eq: eta evolution}), Eq.~\eqref{eqr1} would be modified to~\cite{SM}
\begin{eqnarray}
    \frac{dR_1}{dt} &=& \left[\frac{K_1}{2}-D\right]R_1 -\frac{1}{2}\left[\frac{K^2_1}{K_1+2D}\right]R_1^3 \nonumber\\
    &&+ \frac{D}{ N R_1}\left[1+R_1^2\left(\frac{K_1}{K_1+2D}\right)\right] \nonumber \\
    && +  \sqrt{\frac{D}{N}\left[1-R_1^2\left(\frac{K_1}{K_1+2D}\right)\right]} \xi_r(t).\label{eq: dR1/dt mine}
\end{eqnarray}
We compare Eq.~\eqref{eq: dR1/dt mine} with Eq. (11) of Ref.~\cite{Buendia2025}, which reads as
\begin{eqnarray}
     \frac{dR_1}{dt} &=& \left[\frac{K_1}{2}-D\right]R_1-\frac{K_1}{2}R_1^3+\frac{D\left[1+R_1^2\right]}{NR_1}\nonumber \\
     &&+\sqrt{\frac{D\left[1-R_1^2\right]}{N}}\xi_r(t).\label{eq: dR1/dt Buendia}
\end{eqnarray}
We observe that the coefficient of $R_1^3$ in the two equations does not match. Moreover, inside the brackets of the third and fourth terms on the right hand side, the coefficient of $R_1^2$ does not match. These terms come from the part of $\eta(\theta,\omega,t)$ that nonlinearly depends on $A(t)$ (see the discussions before Eq.~\eqref{eq:CM ansatz}), which in turn originate from the nonlinear $\eta$ contributions in Eq.~\eqref{eq: eta evolution}. Reference~\cite{Buendia2025} used the OA ansatz to deal with these nonlinear $\eta$ contributions, whereas we use center manifold expansion, which is the reason for the mismatch in the corresponding coefficients. Since a center manifold expansion is asymptotically exact near the transition point~\cite{carr1981centre}, our method captures finite-$N$ fluctuations extremely well near the transition point, as compared to Ref.~\cite{Buendia2025}.

\begin{figure}[htbp!]
\includegraphics[width=0.75\linewidth]{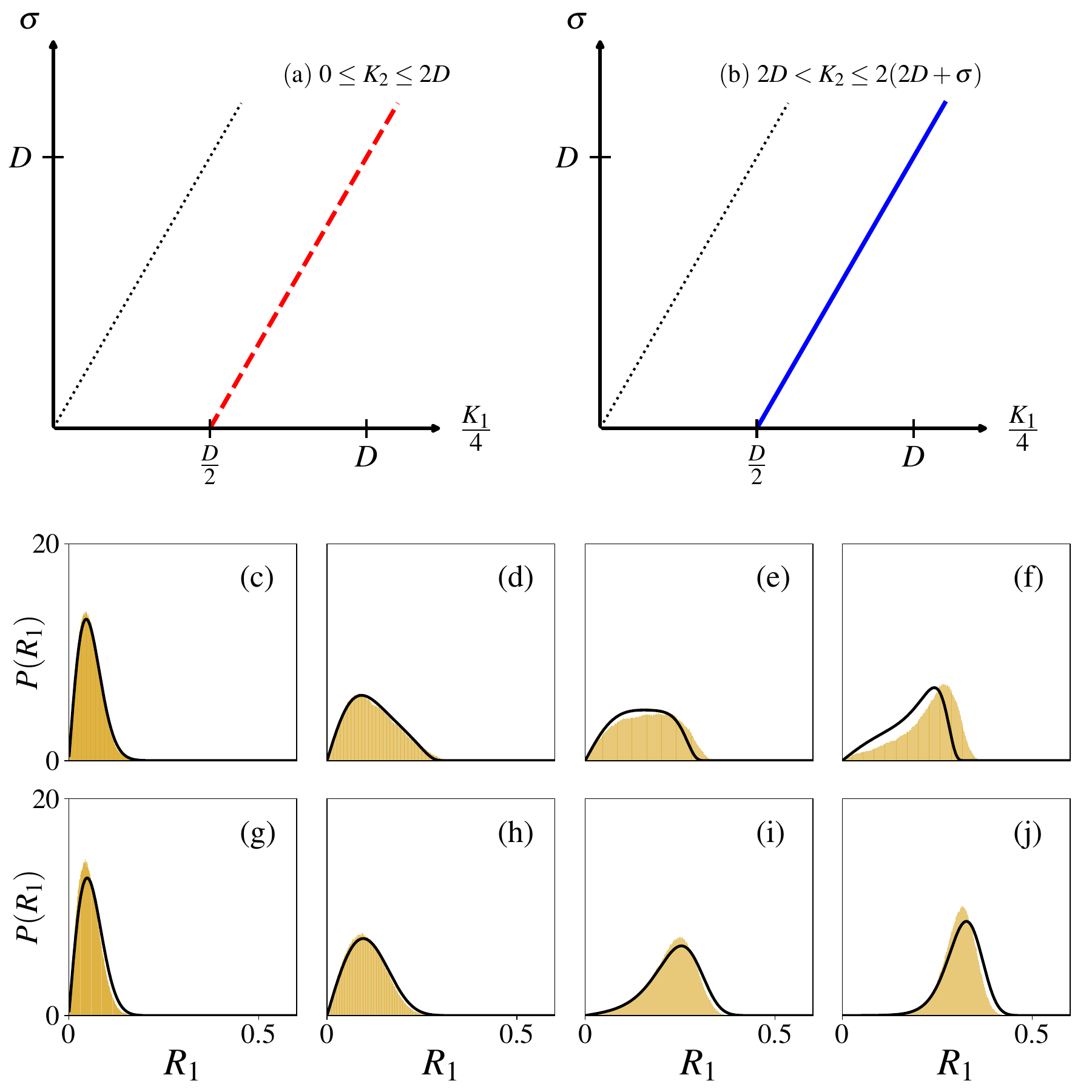}
    \caption{For model in \textit{Application 3}, (a) and (b) show the $N\to \infty$ phase diagram containing the continuous (red dashed line) and first-order (blue solid line) transition lines in the  $\sigma-K_1$ plane for fixed $K_2$. For  $\sigma = 1.0, D =1$, agreement between our theory (lines) and numerical simulation (histogram) for first-order transition ($K_2=2.3$ with $K_1 = 3.9, 3.97, 3.98, 3.99$ and $N =10^4$) is shown is (c) -- (f) and for continuous transition ($K_2=1.0$ with $K_1 = 3.5, 3.9, 4.1, 4.2$ and $N = 2\times 10^3$) is shown is (g) -- (j).}
    \label{fig: 4}
\end{figure}

Equation~\eqref{eq: dR1/dt Buendia}~agrees with Eq.~\eqref{eq: dR1/dt mine} in two limits: (i) When $R_1$ is small, both equations reduce to $d R_1/dt = \left[K_1/2-D\right]R_1+D/(NR_1)+\sqrt{D/N}\xi_r(t)$. For $K_1<2D$ (incoherent phase), when $R_1$ is small,  both the equations describe well finite-size fluctuations in the incoherent phase. (ii) The other limit when Eq.~\eqref{eq: dR1/dt mine} reduces to Eq.~\eqref{eq: dR1/dt Buendia} is when $K_1 \gg D$, enabling us to have $K_1+2D\approx K_1$. Hence, Eq.~\eqref{eq: dR1/dt Buendia} agrees with Eq.~\eqref{eq: dR1/dt mine} when the noise strength $D$ is very small. This reflects the fact that the OA ansatz, on which the analysis in Ref.~\cite{Buendia2025} was based, was proposed for the noiseless Kuramoto model, and only works approximately for very small noise strength. Near the transition point ($K_1 = 2D$), when both $K_1$ and $D$ are of same order, the agreement of~Ref.~\cite{Buendia2025} with simulations suffers.
\begin{center}
    \textbf{\small Appendix E: Results for Application 3}
\end{center}
 The results are shown in Fig.~\ref{fig: 4}.
\begin{center}
    \textbf{\small Appendix F: Fluctuations in $g(\omega)$ --}
\end{center}
\vspace{-0.4cm}
\begin{figure}[htbp!]
\includegraphics[width=0.65\linewidth]{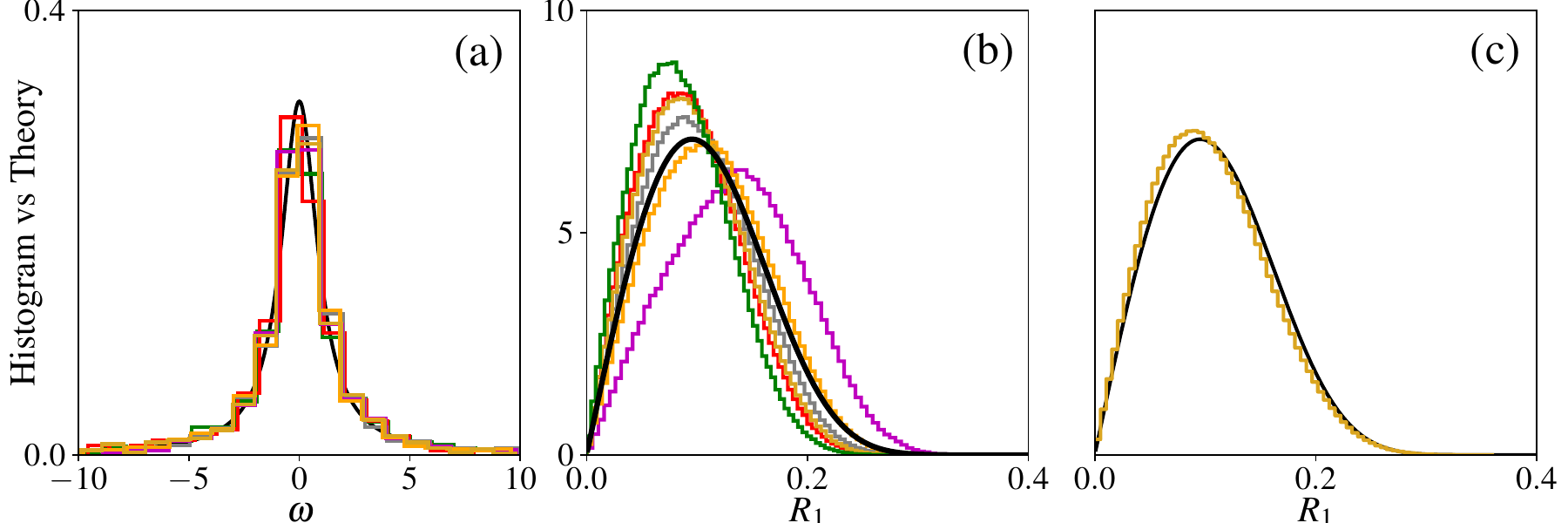}
    \caption{For \textit{Application III} with $K_1=3.9,K_2=1.0,D=1.0, \sigma=1.0, N= 2000$: (a) frequency histograms $\bar{g}_{_N}(\omega)$ from different realizations of sampled frequencies (colored) compared with the Lorentzian $g(\omega)$ (black). These are obtained for typical realizations of the frequencies, where for each oscillator, its frequency is sampled independently from the distribution $g(\omega)$. (b) Steady-state histograms of $R_1$ for the corresponding samples (colored), with theoretical $P(R_1)$ shown in black. (c) Agreement between the average  (colored) of the six histograms presented in (b) and $P(R_1)$ (black).}
    \label{fig: 8}
\end{figure}
For \textit{Application II} we considered the frequency-sampling fluctuations by introducing the parameter $\alpha$ (\textit{Application I} has no frequency disorder). However, for \textit{Application III} involving the heavy-tailed Lorentzian distribution, Fig.~\ref{fig: 8} shows that for small $N$, sample-to-sample fluctuations of $\bar{g}_{_N}(\omega)$ induce visible deviations in the corresponding steady-state histogram of $R_1$ with respect to the distribution of $P(R_1)$ predicted theoretically by replacing $\bar{g}_{_N}(\omega)$ by $g(\omega)$. Nevertheless, averaging these histograms over different frequency realizations gives excellent agreement with $P(R_1)$, demonstrating that our theory captures the mesoscopic fluctuations due to dynamical Gaussian noise, but not the quenched ones, arising from frequency sampling at small $N$. A possible remedy to access the quenched fluctuations is to approximate a given realization $\bar{g}_{_N}(\omega)=N^{-1}\sum_{j=1}^N\delta(\omega-\omega_j)$ by coarse-graining it into $M$ bins within $[-\Omega,\Omega]$, and approximating it as $\bar{g}_{_N}(\omega) \approx \big(\sum_{m=0}^{(M-1)}f_m\big)^{-1}\sum_{m=0}^{(M-1)}f_m\delta[\omega-\Omega(2m+1-M)/M]$, where $f_m$ is the number of sampled frequencies in the $m$-th bin. This amounts to a generalization of our analysis, which we leave for future work.

\clearpage

\onecolumngrid

\setcounter{equation}{0}
\renewcommand{\theequation}{SM\arabic{equation}}

\begin{center}
{\bf Supplementary Information for: ``Finite-size fluctuations for stochastic coupled oscillators: A general theory''}
\end{center}

\tableofcontents

\section{Derivation of Eq.~(5) of the main text \label{sec: sup 1}}

We start with the global noise field $\zeta(\theta,\omega,t)$ in Eq.~(2) of the main text. Since  $\zeta(\theta,\omega,t)$ is $2\pi$-periodic in $\theta$, we can express it in a Fourier series as $\zeta(\theta,\omega,t) = \sum_{l=-\infty}^{+\infty} \xi_l (\omega,t) e^{il \theta}$. The properties of $\zeta(\theta,\omega,t)$  given by $\left \langle \zeta(\theta,\omega,t) \right \rangle = 0$ and $\left \langle \zeta(\theta,\omega,t)\zeta(\theta',\omega',t') \right \rangle = \delta(\theta-\theta')\delta(\omega-\omega')\delta(t-t')$ put further conditions on the Fourier coefficients as
\begin{eqnarray}
 \langle \xi_l(\omega,t)\rangle &=& \langle \xi_l (\omega,t)\xi_m (\omega',t')\rangle =  0,\\
    \langle \xi_l (\omega,t)\xi^{*}_m (\omega',t')\rangle &=& \frac{1}{2\pi}  \delta_{lm}\delta(\omega-\omega')\delta(t-t').
\end{eqnarray}
These relations will be useful later. 

For further computations, we are interested in the bifurcations from the homogeneous stationary state, i.e., the incoherent state. In this parameter region with $N \to \infty$, the density is simply $\bar{F}_{N \to \infty}(\theta, \omega, t \to \infty)= g(\omega)/(2\pi) $, where $g(\omega)$ is the distribution function of the frequencies. When $N$ is finite, we shall assume that it is large enough so that $\bar{g}_N$ can be replaced by $g$. We can then express the finite-$N$ empirical density in the following form:
\begin{eqnarray}
    \bar{F}_N (\theta, \omega, t) = \frac{g(\omega)}{2\pi} + \eta(\theta,\omega,t), \label{eq: F eta relation}
\end{eqnarray}
where $\eta(\theta,\omega,t)$ is small. 

Putting Eq.~\eqref{eq: F eta relation} into Eq.~(2) of the main text, we obtain
\begin{eqnarray}
    \frac{\partial\eta}{\partial t} = \mathcal{L}\eta + \mathcal{N}[\eta]+\sqrt{\frac{2D}{ N}} \frac{\partial}{\partial \theta}\left[\sqrt{\frac{g(\omega)}{2\pi}+\eta} \zeta(\theta,\omega,t)\right], \label{eq: SUP eta evolution0}
\end{eqnarray}
where $\mathcal{L}$ is a linear operator and $\mathcal{N}$ is a non-linear operator, which read as
\begin{eqnarray}
    \mathcal{L}\eta &=&  D\frac{\partial^2 \eta  }{\partial \theta^2} -\omega\frac{\partial \eta }{\partial \theta}  -\frac{g(\omega)}{2\pi}    \int_0^{2\pi} d\theta' \int_{-\infty}^{\infty} d\omega' \partial_\theta f(\theta'-\theta)  \eta(\theta',\omega',t), \label{eq: DK linear}\\
    \mathcal{N}[\eta] &=& -\partial_\theta \left[\eta(\theta,\omega,t)  \int_0^{2\pi} d\theta' \int_{-\infty}^{\infty} d\omega' f(\theta'-\theta)\eta(\theta',\omega',t) \right]. \label{eq: DK non-linear}
\end{eqnarray}
As mentioned in the main text, in the leading order, we may consider the noise term without $\eta$ for further computations, which gives us
\begin{eqnarray}
    \frac{\partial\eta}{\partial t} = \mathcal{L}\eta + \mathcal{N}[\eta]+\sqrt{\frac{Dg(\omega)}{ \pi N}} \frac{\partial}{\partial \theta}\bigg[\zeta(\theta,\omega,t)\bigg]. \label{eq: SUP eta evolution}
\end{eqnarray}
In Eq.~\eqref{eq: SUP eta evolution}, we have kept only the leading order of the noise, which makes the noise term in the equation, which was otherwise multiplicative, to be additive, see Ref.~\cite{Martin2021} for an example where the multiplicative structure is qualitatively important. With $\eta(\theta,\omega,t)$ being small, $\mathcal{L}\eta$ dominates over $\mathcal{N}[\eta]$. Furthermore, $N$ is sufficiently large such that $\mathcal{L}\eta$ also dominates over the noise in Eq.~\eqref{eq: SUP eta evolution}. Hence, at leading order, the dynamics of $\eta$ will be determined by the operator $\mathcal{L}$. 

Before moving forward, let us understand the action of operators $\mathcal{L}$ and $\mathcal{N}$ in detail. Let $\Phi \equiv \Phi(\theta,\omega)$ be a general function. The angle $\theta$ being $2\pi$-periodic, we can use the following Fourier expansions: $\Phi(\theta, \omega) =\sum_{m=-\infty}^{+\infty} \Phi_m(\omega)e^{im\theta}$, $\mathcal{L}\Phi  = \sum_{m=-\infty}^{+\infty} \left (L_m \Phi_m\right)e^{im\theta}$, and $\mathcal{N}\left[ \Phi \right]  = \sum_{m=-\infty}^{+\infty} \mathcal{N}_m\left[ \Phi \right]e^{im\theta}$. Furthermore, we have $f(q) = \sum_{l = 1}^{+\infty} K_l \sin{(lq)}$. Putting these expansions back into Eq.~\eqref{eq: DK linear} and comparing the Fourier modes of both sides, we obtain
\begin{eqnarray}
    L_m \Phi_m = -(m^2D+im\omega)\Phi_m(\omega) +  \frac{g(\omega)}{2}  \int_{-\infty}^{\infty} d\omega' \Phi_m(\omega')\sum_{l=1}^{+\infty} l K_l \left(\delta_{l,m}+\delta_{l,-m}\right), \label{eq: LmPhim}
\end{eqnarray}
where $\delta_{a,b}$'s are Kronecker deltas. Similarly, putting the Fourier expansions back into Eq.~\eqref{eq: DK non-linear}, we obtain
\begin{eqnarray}
    \mathcal{N}_m \left[\Phi\right] = \pi m \sum_{l=1}^{+\infty} K_l \left[ \Phi_{m-l}(\omega) \int_{-\infty}^{+\infty} d\omega' \Phi_l(\omega')-\Phi_{m+l}(\omega) \int_{-\infty}^{+\infty} d\omega' \Phi_{-l}(\omega')\right].  \label{eq: SUP Non Lin etam}
\end{eqnarray}

We now focus on finding the spectrum of the linear operator $\mathcal{L}$; for more details, see \cite{Crawford1994}. Considering $\Psi_m(\theta,\omega) = \psi_m(\omega) e^{im\theta}$ to be its eigenfunction with the eigenvalue $\lambda_m$, the eigenvalue equation $\mathcal{L}\Psi_m = \lambda_m \Psi_m$ gives $L_m \psi_m = \lambda_m \psi_m$. Using Eq.~\eqref{eq: LmPhim}, we obtain
\begin{eqnarray}
\left(\lambda_m+m^2D+im\omega\right)\psi_m(\omega) = \frac{|m| K_{|m|} g(\omega)}{2}  \int_{-\infty}^{\infty} d\omega' \psi_m(\omega').
\label{eq:spec}
\end{eqnarray}
For any $m$ such that $K_{|m|}=0$, the only possible functions $\psi_m$ are singular, corresponding to the existence of a continuous spectrum on the line ${\rm Re}(\lambda)=-Dm^2$. For $m$ such that $K_{|m|}\neq 0$, there is still a continuous spectrum on the line ${\rm Re}(\lambda)=-Dm^2$, but there may also exist nonsingular solutions to \eqref{eq:spec}, corresponding to the discrete spectrum which will be our main interest. They satisfy:
\begin{eqnarray}
\psi_m(\omega) = \frac{|m| K_{|m|}g(\omega)}{2\left(\lambda_m+m^2D+im\omega\right)}  \int_{-\infty}^{\infty} d\omega' \psi_m(\omega'). \label{eq: Psim self con}
\end{eqnarray}
Integrating both sides of Eq.~\eqref{eq: Psim self con} with respect to $\omega$ and noting that $\int_{-\infty}^{\infty} d\omega' \psi_m(\omega') \neq 0$, we obtain the secular equation determining the eigenvalues, which reads as $\Lambda_m(\lambda_m)=0$, where $\Lambda_m$ is the spectral function
\begin{eqnarray}
    \Lambda_m(x) = 1-\frac{|m| K_{|m|}}{2}\int_{-\infty}^{\infty} d\omega' \frac{ g(\omega')}{\left(x+m^2D+im\omega'\right)}. \label{eq: Spectral L}
\end{eqnarray}
Interestingly, the spectral function satisfies the identity $\left[\Lambda_m(x)\right]^{*} = \Lambda_{-m}(x^{*})$, with the star denoting complex conjugation. Hence, if $\lambda_m$ is a root of $\Lambda_m(x)$, then $\lambda^{*}_m$ is also a root of $\Lambda_{-m}(x)$. Clearly, the eigenvalues $\lambda_m$ change upon changing the interaction strength $K_{|m|}$ of the corresponding Fourier mode. Using the normalization $\int_{-\infty}^{\infty} d\omega' \psi_m(\omega') = 1$, we can write
\begin{eqnarray}
\psi_m(\omega) = \frac{|m| K_{|m|} g(\omega)}{2\left(\lambda_m+m^2D+im\omega\right)}  . \label{eq: Psim sup}
\end{eqnarray}

We may further define an adjoint operator for the linear operator $\mathcal{L}$ from the definition $\left(A , \mathcal{L}B \right)=\left(\mathcal{L}^\dagger A , B \right)$, where the inner product is given by $\left( A,B\right) = \int_0^{2\pi}\int_{-\infty}^{+\infty} d\theta' d\omega' A^{*}(\theta',\omega')B(\theta',\omega')$. Defining the Fourier expansions $A(\theta, \omega) \equiv\left(2\pi\right)^{-1}\sum_{m=-\infty}^{+\infty} A_m(\omega)e^{im\theta}$ and $\mathcal{L}^\dagger A\equiv\left(2\pi\right)^{-1} \sum_{m=-\infty}^{+\infty} \left(L^\dagger_m A_m\right)e^{im\theta}$ and using the definition of $\mathcal{L}^\dagger$ along with Eq.~\eqref{eq: LmPhim}, we obtain
\begin{eqnarray}
    L^\dagger_m A_m = -(m^2D-im\omega)A_m(\omega) +  \frac{1}{2}  \int_{-\infty}^{\infty} d\omega' g(\omega')A_m(\omega')\sum_{l=1}^{+\infty} l K_l \left(\delta_{l,m}+\delta_{l,-m}\right). \label{eq: Ldag_mPhim}
\end{eqnarray}
The spectrum of $\mathcal{L}^\dagger$ may be found in the following way. Considering $\tilde{\Psi}_m(\theta,\omega) = \left(2\pi\right)^{-1}\tilde{\psi}_m(\omega) e^{im\theta}$ to be its eigenfunction with the eigenvalue $\lambda_{\dagger,m}$, the eigenvalue equation $\mathcal{L}^\dagger\tilde{\Psi}_m = \lambda_{\dagger,m} \tilde{\Psi}_m$ gives $L^\dagger_m \tilde{\psi}_m = \lambda_{\dagger,m} \tilde{\psi}_m$. Using Eq.~\eqref{eq: Ldag_mPhim} , we obtain
\begin{eqnarray}
\left(\lambda_{\dagger,m}+m^2D-im\omega\right)\tilde{\psi}_m(\omega) = \frac{|m| K_{|m|}}{2}  \int_{-\infty}^{\infty} d\omega' g(\omega')\tilde{\psi}_m(\omega').
\end{eqnarray}
Similar to the case of $\mathcal{L}$, if there exists an $m$ such that $K_{|m|}=0$, then we have the trivial solution $\lambda_{\dagger,m} = -m^2D+im\omega$ with $\tilde{\psi}_m(\omega') = \delta(\omega'-\omega)$. On the contrary, if $K_{|m|}\neq0$, we have the nontrivial solution
\begin{eqnarray}
\tilde{\psi}_m(\omega) = \frac{|m| K_{|m|}}{2\left(\lambda_{\dagger,m}+m^2D-im\omega\right)}  \int_{-\infty}^{\infty} d\omega' g(\omega')\tilde{\psi}_m(\omega'). \label{eq: PsiTilde_m self con}
\end{eqnarray}
Multiplying both sides of Eq.~\eqref{eq: PsiTilde_m self con} by $g(\omega)$ and integrating with respect to frequency, we obtain that the eigenvalue $\lambda_{\dagger,m}$ of the operator $\mathcal{L}^\dagger$ is the root of the spectral function
\begin{eqnarray}
    \Lambda_{\dagger,m}(x) = 1-\frac{|m| K_{|m|}}{2}\int_{-\infty}^{\infty} d\omega' \frac{ g(\omega')}{\left(x+m^2D-im\omega'\right)}. \label{eq: Spectral L dagger}
\end{eqnarray}
Comparing Eqs.~\eqref{eq: Spectral L} and \eqref{eq: Spectral L dagger}, we obtain $ \Lambda_{\dagger,m}(x)  = \Lambda_{-m}(x) $. Hence, the eigenvalues of $\mathcal{L}^\dagger$ corresponding to $m$-th Fourier mode are exactly the same as the eigenvalues of $\mathcal{L}$ corresponding to $(-m)$-th Fourier mode. These in turn are the same as the complex conjugates of the eigenvalues of $\mathcal{L}$ corresponding to $m$-th Fourier mode. Mathematically, $\lambda_{\dagger,m} = \lambda_{-m} = \lambda^{*}_m.$ Furthermore, imposing the the orthonormality condition $\left(\tilde{\Psi}_m,\Psi_{m'}\right) = \delta_{m,m'}$ gives
\begin{eqnarray}
    \tilde{\psi}_{m}(\omega) = \frac{1}{\left[\Lambda'_m (\lambda_{m})\right]^{*} }\frac{1}{(\lambda_{\dagger,m}+m^2D-im\omega)} . \label{eq: SUP psi tilde}
\end{eqnarray}

In the incoherent phase, the real part of all the eigenvalues are negative, i.e.,  $\Re\left(\lambda_m\right) <0$. Hence, any perturbation of the incoherent state will die down fast in time, making it a stable state. Upon changing $K_{|m|}$ for a particular $m$, keeping all other interaction strengths constant, it may so happen that once $K_{|m|}$ crosses a particular value, say $K_{|m|}^\mathrm{c}$, $\Re\left(\lambda_m\right)$ and $\Re\left(\lambda_{-m}\right)$ for that particular $m$ change sign and become positive. As a result, any perturbation along the direction of $\Psi_{m}(\theta,\omega)$ and $\Psi_{-m}(\theta,\omega)$ will grow with time, making the incoherent phase unstable. At linear order in the perturbation, the dynamics is then confined to the linear subspace ${\rm Span}(\Psi_m,\Psi_{-m})$. In the deterministic case (i.e., without noise) and close to the bifurcation/transition, the nonlinearities drive the dynamics outside ${\rm Span}(\Psi_m,\Psi_{-m})$; however, it remains confined to a manifold (the center manifold), which is a nonlinear deformation of ${\rm Span}(\Psi_m,\Psi_{-m})$. The goal of the (deterministic) center manifold expansion is to determine at the same time this manifold and the slow dynamics that takes place on it (see Refs.~\cite{Strogatz1990, Strogatz1992, Crawford1994, Crawford1999, Gupta2018, PhysRevResearch.2.023183} for implementations in the context of synchronization models in the thermodynamic limit). Our crucial hypothesis is that the finite-$N$ noise will not drive the system too far from the deterministic center manifold, so that we only have to understand how the noise impacts the dynamics on the center manifold. Similar effect will also be there when $\Re(\lambda_{\pm m})$ is least negative among the rest of the eigenvalues. In this case also, there will be a timescale separation of the dynamics of $\eta(\theta,\omega,t)$ on the subspace ${\rm Span}(\Psi_m,\Psi_{-m})$ and orthogonal to it, resulting in a net dynamics of $\eta(\theta,\omega,t)$ on the center manifold.

For the next part of the calculation, we will assume that $K_1$ is varied keeping all the other $K_{|m|}$'s fixed. Hence, $\Re\left(\lambda_{\pm1}\right)$ changes sign, while all other $m$'s have $\Re\left(\lambda_m\right)<0$. As a result, the Kuramoto-Daido order parameter $R_1$ shows a phase transition. This calculation may be reproduced for any other order parameter $R_m$ showing a phase transition.

Since, $\Re\left(\lambda_{\pm1}\right)$ is changing sign, let us assume that we are at a parameter region such that $\Re\left(\lambda_{\pm1}\right)$ is either positive or least negative among all other $\Re\left(\lambda_m\right)$. Hence, following the previous discussion, we may write
\begin{eqnarray}
    \eta(\theta,\omega,t) = A(t) \Psi_1(\theta,\omega) + A^{*}(t) \Psi_{-1}(\theta,\omega) + S(\theta,\omega,t).
\end{eqnarray}
Here $S(\theta,\omega,t)$ incorporates (i) the deformation of the subspace
${\rm Span}(\Psi_m,\Psi_{-m})$ into the center manifold, due to the nonlinearities; (ii) the effect of the noise, due to finite-size effects. According to the above discussion, we shall decouple these two effects, assuming that the noise is small enough so that it does not modify much the computation of the center manifold. 
We may assume that $S(\theta,\omega,t)$ is ``orthogonal" to the space spanned by $\Psi_{1}(\theta,\omega)$ and $\Psi_{-1}(\theta,\omega)$ (in the sense that $\left(\tilde{\Psi}_{\pm 1}, S\right)=0$, where $(,)$ denotes inner product as defined previously). Next, we use the usual center manifold ansatz~\cite{Crawford1994}, which assumes that $S(\theta,\omega,t)$ may be written as a function of the amplitudes along $\Psi_{\pm1}(\theta,\omega)$, i.e., $S(\theta,\omega,t) = W\left[A,A^{*}\right]$. Hence, we may write
\begin{eqnarray}
    \eta(\theta,\omega,t) = A(t) \Psi_1(\theta,\omega) + A^{*}(t) \Psi_{-1}(\theta,\omega) + W\left[A,A^{*}\right]. \label{eq: SUP eta CM}
\end{eqnarray}
The orthogonality condition gives $\left(\tilde{\Psi}_{\pm1},W\left[A,A^{*}\right]\right) = 0$. Taking the derivative with respect to time, we immediately get $\left(\tilde{\Psi}_{\pm1},\partial W/\partial t\right) = 0$. Now, $\eta(\theta,\omega,t)$ is $2\pi$-periodic in $\theta$. Hence, we may expand it in a Fourier series as $\eta(\theta,\omega,t) = \sum_{m=-\infty}^{+\infty} \eta_m(\omega,t)e^{im\theta}$. Normalization of $\bar{F}_N (\theta, \omega, t)$ immediately gives $\eta_0(\omega,t) = 0$. Moreover, since $\eta(\theta,\omega,t)$ is $2\pi$-periodic, $W\left[A,A^{*}\right]$ also becomes $2\pi$-periodic following Eq.~\eqref{eq: SUP eta CM}. Hence, we may expand it in a Fourier series as $W\left[A,A^{*}\right] = \sum_{m=-\infty}^{+\infty} W_m\left[A,A^{*}\right]e^{im\theta}$. Comparing the Fourier coefficients on both sides of Eq.~\eqref{eq: SUP eta CM}, we obtain
\begin{subequations} \label{eq:eta sup}
\begin{align}
    &\eta_0(\omega,t) = W_0\left[A,A^{*}\right] = 0, \label{eq:eta0} \\
   &\eta_1(\omega,t) = A(t)\psi_1(\omega)+W_1\left[A,A^{*}\right] , \label{eq:eta1} \\
&\eta_{-1}(\omega,t) = A^{*}(t)\psi_{-1}(\omega)+W_{-1}\left[A,A^{*}\right] , \label{eq:eta m1}\\
&\eta_{m}(\omega,t) = W_{m}\left[A,A^{*}\right]~\forall~|m|>1.\label{eq:eta m}
\end{align}
\end{subequations}

To find the mathematical form of the above Fourier coefficients, we use the rotational symmetry of the system given by Eq. (1) of the main text. If all the oscillator phases are rotated by the same angle, i.e., under the transformation $\theta_j \to \theta_j+\alpha~\forall~j,~\forall~\alpha$, Eq. (1) of the main text remains invariant. Hence, the density $\eta(\theta,\omega,t)$ also remains invariant under this transformation. Under this transformation, clearly, the eigenfunctions of the operator $\mathcal{L}$ transform as $\Psi_m(\theta,\omega)\to \Psi_m(\theta,\omega)e^{im\alpha}$. Hence, to keep $\eta(\theta,\omega,t)$ invariant, $A(t)$ should transform as $A(t)\to A(t)e^{-i\alpha}$, while $A^{*}(t)$ should transform as $A^{*}(t)\to A^{*}(t)e^{i\alpha}$, and $W_m\left[A,A^{*}\right]$ should transform as $W_m\left[A,A^{*}\right] \to W_m\left[A,A^{*}\right] e^{-im\alpha}$. Since, $W_m\left[A,A^{*}\right]$ is a function of only $A$ and $A^{*}$, to preserve the transformation structure, it must have the form $W_m\left[A,A^{*}\right] = A^{m} \mathbb{W}_m\left(|A|^2\right)$, where $\mathbb{W}_m\left(|A|^2\right)$ may be written as $\mathbb{W}_m\left(|A|^2\right) = \mathrm{w}_{m,0} + \mathrm{w}_{m,2} |A|^2 + \mathrm{w}_{m,4} |A|^4 + \ldots$. The center manifold is tangent to the subspace ${\rm Span}(\Psi_1,\Psi_{-1})$ at $A=A^\ast=0$; this imposes $W[0,0] = \left.\frac{\partial W}{\partial A}\right|_{A=A^{*}=0}= \left.\frac{\partial W}{\partial A^{*}}\right|_{A=A^{*}=0} =0$, which gives $\mathrm{w}_{1,0}=0$. Combining all these pieces of information, we get
\begin{subequations} \label{eq:W}
\begin{align}
   & W_0\left[A,A^{*}\right]= 0, \label{eq:W0} \\
    &W_1\left[A,A^{*}\right]= A\left|A\right|^2 \left(\mathrm{w}_{1,2}+\mathrm{w}_{1,4} |A|^2 + \ldots\right), \label{eq:W1} \\
    &W_m\left[A,A^{*}\right] = A^m \left(\mathrm{w}_{m,0}+\mathrm{w}_{m,2} |A|^2+\mathrm{w}_{m,4} |A|^4 + \ldots\right)~\forall~ m>1, \label{eq:Wm}\\
    &W_{-m}\left[A,A^{*}\right] = \left( W_m\left[A,A^{*}\right]\right)^{*}~\forall~m\neq0. \label{eq:Wmneg}
\end{align}
\end{subequations}
Equations~\eqref{eq:eta0}~--~\eqref{eq:eta m}~and~\eqref{eq:W1} -- \eqref{eq:Wmneg} may be written in a compact form as follows:
\begin{eqnarray}
     \eta_{m}(\omega,t)= A^m\sum_{p=0}^{+\infty} w_{m,2p}|A|^{2p};~\forall~m \geq 0,
\end{eqnarray}
where $w_{0,2p}=0~\forall~p$ and $w_{1,0} = \psi_{1}(\omega)$ and $w_{m,2p}=\mathrm{w}_{m,2p}$ for any other combination of $m>0$ and $p$, and with
\begin{eqnarray}
     \eta_{-m}(\omega,t)= \left[\eta_{m}(\omega,t)\right]^{*}.
\end{eqnarray}
We remark that from the definition of the order parameter, we have
\begin{eqnarray}
    Z_1 = R_1e^{i\psi_1} &=& \int_0^{2\pi}d\theta'e^{i\theta'}\int_{-\infty}^{+\infty}d\omega'\eta(\theta',\omega',t) = 2\pi \int_{-\infty}^{+\infty}d\omega'\eta_{-1}(\omega',t) \nonumber \\
    &=&2\pi A^{*}\left[1+\sum_{n=1}^{+\infty} \mathcal{A}_{1,2n}^{*}|A|^{2n}\right] = 2\pi A^{*}+\mathcal{O}\left(A^{*}|A|^2\right),
\end{eqnarray}
where $\mathcal{A}_{m,2p} \equiv \int_{-\infty}^{\infty} d\omega ~w_{m,2p}$. Hence, calculating the evolution equation of the amplitude $A(t)$ will give us the finite-size fluctuations in the order parameter $R_1(t)$.

We now turn to the dynamics to determine both the $w_{m,2p}$'s and the reduced dynamics for $A,A^\ast$. We start by taking the time derivative on the both side of Eq.~\eqref{eq: SUP eta CM}, which gives
\begin{eqnarray}
    \frac{\partial \eta(\theta,\omega,t)}{\partial t} = \dot{A}(t) \Psi_1(\theta,\omega) + \dot{A}^{*}(t) \Psi_{-1}(\theta,\omega) + \frac{\partial W\left[A,A^{*}\right]}{\partial t}, \label{eq: SUP ddt eta CM}
\end{eqnarray}
where the dot represents derivative with respect to time. We now compute the inner product on both side of Eq.~\eqref{eq: SUP ddt eta CM} with $\tilde{\Psi}_1(\theta,\omega)$. Using the orthonormality properties of $\Psi_m(\theta,\omega)$ and $\tilde{\Psi}_m(\theta,\omega)$ along with the condition on $W\left[A,A^{*}\right]$, we obtain
\begin{eqnarray}
    \dot{A}(t) = \left( \tilde{\Psi}_1 , \frac{\partial \eta}{\partial t}\right). \label{eq: SUP ddt A}
\end{eqnarray}
Using Eq.~\eqref{eq: SUP eta evolution} in Eq.~\eqref{eq: SUP ddt A}, we obtain
\begin{eqnarray}
    \dot{A}(t) = \left( \tilde{\Psi}_1 ,\mathcal{L}\eta \right)+\left( \tilde{\Psi}_1 , \mathcal{N}[\eta]\right)+\left( \tilde{\Psi}_1 ,\sqrt{\frac{Dg(\omega)}{\pi N}} \frac{\partial \zeta(\theta,\omega,t) }{\partial \theta}\right). \label{eq: SUP ddt A detailed}
\end{eqnarray}
The first term on the right hand side of Eq.~\eqref{eq: SUP ddt A detailed} gives
\begin{eqnarray}
    \left( \tilde{\Psi}_1 ,\mathcal{L}\eta \right) = \left(\mathcal{L}^\dagger \tilde{\Psi}_1 ,\eta \right) = \left(\lambda_{\dagger,1} \tilde{\Psi}_1 ,\eta \right) =  \lambda^{*}_{\dagger,1}\left( \tilde{\Psi}_1 ,\eta \right) = \lambda_1 A(t). \label{eq: SUP ddt A detailed RHS1}
\end{eqnarray}
The second term on the right hand side of Eq.~\eqref{eq: SUP ddt A detailed} gives
\begin{eqnarray}
    \left(\tilde{\Psi}_1 , \mathcal{N}[\eta]\right) =  \int_{-\infty}^{+\infty} d\omega' \tilde{\psi}^{*}_{1}(\omega')\mathcal{N}_1[\eta]\label{eq: SUP ddt A detailed RHS2}.
\end{eqnarray}
The third term on the right hand side of Eq.~\eqref{eq: SUP ddt A detailed} gives
\begin{eqnarray}
    \left( \tilde{\Psi}_1 ,\sqrt{\frac{Dg(\omega)}{\pi N}} \frac{\partial \zeta(\theta,\omega,t) }{\partial \theta}\right) = i \sqrt{\frac{D}{\pi N}} \int_{-\infty}^{+\infty} d\omega'\tilde{\psi}^{*}_{1}(\omega')\sqrt{g(\omega')}  \xi_1(\omega',t)\equiv \mathbb{O}(t)\label{eq: SUP ddt A detailed RHS3}.
\end{eqnarray}
Clearly, $\left \langle \mathbb{O}(t)\right\rangle = \left \langle \mathbb{O}(t)\mathbb{O}(t')\right\rangle=0$ and $\left \langle \mathbb{O}(t)\mathbb{O}^{*}(t')\right\rangle = D_\mathrm{eff}\left(2\pi^2 N \right)^{-1} \delta(t-t'')$, where $D_\mathrm{eff} = D\int_{-\infty}^{+\infty} d\omega'\left|\tilde{\psi}_{1}(\omega')\right|^2g(\omega')$. We may then replace the noise term $\mathbb{O}(t)$ by $\sqrt{D_\mathrm{eff}\left(2\pi^2 N \right)^{-1}}\xi(t)$ with $\left \langle \xi(t)\right\rangle = \left \langle \xi(t)\xi(t')\right\rangle=0$ and $\left \langle \xi(t)\xi^{*}(t')\right\rangle = \delta(t-t')$. Putting all of these together along with Eqs.~\eqref{eq: SUP ddt A detailed RHS1} and \eqref{eq: SUP ddt A detailed RHS2} into Eq.~\eqref{eq: SUP ddt A detailed}, we obtain
\begin{eqnarray}
    \dot{A}(t) = \lambda_1 A(t)+\int_{-\infty}^{+\infty} d\omega' \tilde{\psi}^{*}_{1}(\omega')\mathcal{N}_1[\eta]+\sqrt{\frac{D_\mathrm{eff}}{2\pi^2N}}\xi(t) \label{eq: SUP ddt A almost finished}.
\end{eqnarray}

We now focus on the second term on the right hand side of Eq.~\eqref{eq: SUP ddt A almost finished}. Using the expression of $\mathcal{N}_1[\eta]$ from Eq.~\eqref{eq: SUP Non Lin etam} along with Eqs.~\eqref{eq:eta sup} and~\eqref{eq:W}, we may expand $\mathcal{N}_1[\eta]$ in terms of $A$ and $A^{*}$; It reads as
\begin{eqnarray}
    \mathcal{N}_1[\eta] =&& -A\left|A\right|^2\pi\left[K_1 w_{2,0}-K_2\psi_{-1} \int_{-\infty}^{+\infty} d\omega' w_{2,0}\right] \nonumber\\
    &&-A\left|A\right|^4\pi
    \left[K_1 \left( w_{2,2}+w_{2,0} \int_{-\infty}^{+\infty} d\omega' w^{*}_{1,2}\right)
    -K_2 \left(\psi_{-1} \int_{-\infty}^{+\infty} d\omega' w_{2,2}+w^{*}_{1,2}\int_{-\infty}^{+\infty} d\omega'w_{2,0}-w_{3,0}\int_{-\infty}^{+\infty} d\omega'w^{*}_{2,0}\right)\right. \nonumber\\
    &&\hspace{2.5cm}\left.-K_3 w^{*}_{2,0} \int_{-\infty}^{+\infty} d\omega'w_{3,0}\right] \nonumber\\
    && -A |A|^6 \pi \Bigg[ 
qK_1 \left( w_{2,4} + w_{2,2} \int_{-\infty}^{+\infty} d\omega'\, w^{*}_{1,2} + w_{2,0} \int_{-\infty}^{+\infty} d\omega'\, w^{*}_{1,4} \right) \nonumber \\
&& \quad - K_2 \left( 
\psi_{-1} \int_{-\infty}^{+\infty} d\omega'\, w_{2,4} + 
w^{*}_{1,2} \int_{-\infty}^{+\infty} d\omega'\, w_{2,2} + 
w^{*}_{1,4} \int_{-\infty}^{+\infty} d\omega'\, w_{2,0} - 
w_{3,0} \int_{-\infty}^{+\infty} d\omega'\, w^{*}_{2,2} - 
w_{3,2} \int_{-\infty}^{+\infty} d\omega'\, w^{*}_{2,0}
\right) \nonumber \\
&& \quad - K_3 \left(
w^{*}_{2,2} \int_{-\infty}^{+\infty} d\omega'\, w_{3,0} + 
w^{*}_{2,0} \int_{-\infty}^{+\infty} d\omega'\, w_{3,2} - 
w_{4,0} \int_{-\infty}^{+\infty} d\omega'\, w^{*}_{3,0}
\right) \nonumber \\
&& \quad - K_4 w^{*}_{3,0} \int_{-\infty}^{+\infty} d\omega'\, w_{4,0}
\Bigg] + \cdots,
\end{eqnarray}
where $w_{1,2},w_{2,0},w_{2,2},w_{3,0}$ and $\psi_{-1}$ are functions of $\omega$. Interestingly, the coefficient of $A|A|^{2b}$ only depends on $K_1, \ldots,K_{b+1}$. Using Eq.~\eqref{eq: SUP ddt A almost finished}, we obtain
\begin{eqnarray}
    \int_{-\infty}^{+\infty} d\omega' \tilde{\psi}^{*}_{1}(\omega')\mathcal{N}_1[\eta] = -c_3 A|A|^2 -c_5 A|A|^4 + \cdots,
\end{eqnarray}
where we have
\begin{eqnarray}
    \hspace{-0.6cm}c_3 \equiv && \pi\left[K_1\int_{-\infty}^{+\infty} d\omega' \tilde{\psi}^{*}_{1} w_{2,0}-K_2\int_{-\infty}^{+\infty} d\omega' \tilde{\psi}^{*}_{1}\psi_{-1} \int_{-\infty}^{+\infty} d\omega' w_{2,0}\right], \label{eq: SUP C3 EXP}\\
    \hspace{-0.6cm}c_5 \equiv && \pi
    \left[K_1 \left( \int_{-\infty}^{+\infty} d\omega' \tilde{\psi}^{*}_{1} w_{2,2}+\int_{-\infty}^{+\infty} d\omega' \tilde{\psi}^{*}_{1} w_{2,0} \int_{-\infty}^{+\infty} d\omega' w^{*}_{1,2}\right)
    -K_2 \left(\int_{-\infty}^{+\infty} d\omega' \tilde{\psi}^{*}_{1} \psi_{-1} \int_{-\infty}^{+\infty} d\omega' w_{2,2}\right.\right. \nonumber\\
   \hspace{-0.6cm} &&\left.\left.+\int_{-\infty}^{+\infty} d\omega' \tilde{\psi}^{*}_{1} w^{*}_{1,2}\int_{-\infty}^{+\infty} d\omega'w_{2,0}-\int_{-\infty}^{+\infty} d\omega' \tilde{\psi}^{*}_{1}w_{3,0}\int_{-\infty}^{+\infty} d\omega'w^{*}_{2,0}\right)-K_3 \int_{-\infty}^{+\infty} d\omega' \tilde{\psi}^{*}_{1}w^{*}_{2,0} \int_{-\infty}^{+\infty} d\omega'w_{3,0}\right].\label{eq: SUP C5 EXP}
\end{eqnarray}

Now, we have to find the expression of $w_{m,2p}$'s. Here, we sketch the method for finding $w_{2,0}$. The rest of the computation for other $w_{m,2p}$'s may be pursued in a similar way. We start from Eq.~\eqref{eq: SUP ddt eta CM}. Combining it with Eqs.~\eqref{eq: SUP eta evolution} and~\eqref{eq: SUP eta CM}, we obtain
\begin{eqnarray}
    \frac{\partial W}{\partial t} = \mathcal{L}W + \mathcal{N}\left[\eta\right] + i\sqrt{\frac{D g(\omega)}{\pi N}} \sum_{m=-\infty}^{+\infty}m~ \xi_m(\omega,t)e^{im\theta}-\left[\left(\dot{A}-\lambda_1 A\right)\Psi_{1}(\theta,\omega)+\mathrm{c.c.}\right], \label{eq: SUP dWdt}
\end{eqnarray}
where $\mathrm{c.c.}$ means complex conjugate. Comparing the second Fourier mode of both sides of Eq.~\eqref{eq: SUP dWdt}, we obtain
\begin{eqnarray}
    \frac{dW_2}{dt} = L_2W_2 + \mathcal{N}_2\left[\eta\right] + i\sqrt{\frac{4D~g(\omega)}{\pi N}}   \xi_2(\omega,t). \label{eq: SUP dW2dt}
\end{eqnarray}
The left hand side of Eq.~\eqref{eq: SUP dW2dt} to leading order gives: $dW_2/dt \approx 2 A\dot{A}w_{2,0} \approx 2\lambda_1A^2w_{2,0}$. Using Eq.~\eqref{eq: LmPhim}, the first term on the right hand side of Eq.~\eqref{eq: SUP dW2dt} to leading order gives $L_2W_2 \approx -(4D+i2\omega)A^2w_{2,0} + K_2 g(\omega)  A^2 \int_{-\infty}^{\infty} d\omega' w_{2,0}$. Similarly, using Eq.~\eqref{eq: SUP Non Lin etam}, the second term on the right hand side of Eq.~\eqref{eq: SUP dW2dt} to leading order gives $\mathcal{N}_2\left[\eta\right] \approx 2\pi K_1 A^2\psi_1(\omega) $. Furthermore, if $N$ is large enough, we may ignore the third term on the right hand side of Eq.~\eqref{eq: SUP dW2dt}. Combining all of these, we finally obtain
\begin{eqnarray}
    w_{2,0} = \frac{\pi K_1 \psi_1(\omega)   }{\left(\lambda_1 + 2D+i\omega\right)}+\frac{K_2 g(\omega)   }{2\left(\lambda_1 + 2D+i\omega\right)}\int_{-\infty}^{\infty} d\omega' w_{2,0}.
\end{eqnarray}
Following similar steps, we find the rest of the $w_{a,b}$'s, which read as
\begin{eqnarray}
    w_{3,0} &=& \frac{\pi \left[K_1w_{2,0}+K_2 \psi_1(\omega)\int_{-\infty}^{\infty} d\omega' w_{2,0}\right]   }{\left(\lambda_1 + 3D+i\omega\right)}+\frac{K_3 g(\omega)   }{2\left(\lambda_1 + 3D+i\omega\right)}\int_{-\infty}^{\infty} d\omega' w_{3,0},
\end{eqnarray}
and
\begin{eqnarray}
w_{1,2} =&& \frac{c_3 \psi_1(\omega)-\pi \left[K_1w_{2,0}-K_2 \psi_{-1}(\omega)\int_{-\infty}^{\infty} d\omega' w_{2,0}\right]   }{\left(2\lambda_1+\lambda_1^{*} + D+i\omega\right)}+\frac{K_1g(\omega)   }{2\left(2\lambda_1+\lambda_1^{*} + D+i\omega\right)}\int_{-\infty}^{\infty} d\omega' w_{1,2},\\
    w_{2,2} =&& \frac{2c_3w_{2,0}+2\pi K_1\left[w_{1,2} + \psi_1(\omega)\int_{_\infty}^{+\infty }d\omega' w_{1,2}-w_{3,0}\right] +2\pi K_3 \psi_{-1}(\omega)  \int_{_\infty}^{+\infty }d\omega' w_{3,0}  }{\left(3\lambda_1 +\lambda^{*}_1+ 4D+i2\omega\right)}\nonumber\\
    &&+\frac{K_2 g(\omega)   }{\left(3\lambda_1 +\lambda^{*}_1+ 4D+i2\omega\right)}\int_{-\infty}^{\infty} d\omega' w_{2,2}.
\end{eqnarray}

We observe that all the $w_{m,2p}$'s computed till now have the following self-consistent form
\begin{eqnarray}
    w_{m,2p} = a^{(m,2p)}(w)+b^{(m,2p)}(\omega)\int_{-\infty}^{\infty} d\omega'~w_{m,2p}. \label{eq: wm2p explain 0}
\end{eqnarray}
From these equations, to obtain an expression of $w_{m,2p}$, we first integrate Eq.~\eqref{eq: wm2p explain 0} with respect to $\omega$ and rearrange it to obtain an expression of $\int_{-\infty}^{\infty} d\omega' w_{m,2p}$. Putting it back into Eq.~\eqref{eq: wm2p explain 0}, we finally obtain
\begin{eqnarray}
    w_{m,2p} = a^{(m,2p)}(w) + b^{(m,2p)}(\omega) \frac{\int_{-\infty}^{\infty} d\omega'~a^{(m,2p)}(w')  }{1-\int_{-\infty}^{\infty} d\omega'~b^{(m,2p)}(w') }.
\end{eqnarray}

We have now calculated all relevant quantities. 
Combining everything we have, we may express Eq.~\eqref{eq: SUP ddt A almost finished} as
\begin{eqnarray}
    \dot{A}(t) = \lambda_1 A(t)-c_3 A|A|^2 -c_5 A|A|^4 +\sqrt{\frac{D_\mathrm{eff}}{2\pi^2N}}\xi(t) \label{eq: SUP ddt A finished}.
\end{eqnarray}

Proceeding as above, we obtain in general that
\begin{eqnarray}
    \dot{A}(t) = \lambda_1 A(t)-\sum_{n=1}^{\infty}c_{2n+1}A|A|^{2n}+\sqrt{\frac{D_\mathrm{eff}}{2\pi^2N}}\xi(t),\label{eq:reduced sup}
\end{eqnarray}
where we have
\begin{eqnarray}
   c_{2n+1} &=& \pi \sum_{l=1}^{\infty} K_l\sum_{q=0}^{n-l+1}\left[\mathcal{B}_{l+1,2q}\mathcal{A}^{*}_{l,2(n-l-q)} \Theta(n-l-q)-\mathcal{C}^{*}_{l-1,2q}\mathcal{A}_{l,2(n-l+1-q)}\right], \label{eq: c 2n+1 general}
\end{eqnarray}
for $n=1,2,\ldots$ and $\Theta(x) = 1~\forall~x\geq0$ and $\Theta(x) = 0~\forall~x<0$, and with
\begin{eqnarray}
    \mathcal{A}_{m,2p} &&\equiv \int_{-\infty}^{\infty} d\omega ~w_{m,2p},\\
    \mathcal{B}_{m,2p} &&\equiv \int_{-\infty}^{\infty} d\omega ~\tilde{\psi}^{*}_1~w_{m,2p},\\
    \mathcal{C}_{m,2p} &&\equiv \int_{-\infty}^{\infty} d\omega ~\tilde{\psi}_1~w_{m,2p},
\end{eqnarray}
where we have
\begin{eqnarray}
    \nonumber w_{m,2p} =&&\left.\left[ \frac{1}{(m+p)\lambda_1+p\lambda_1^{*}+m^2D+im\omega}\right] \right[  \sum_{q=0}^{p-1}\left[(m+q)c_{2(p-q)+1}+qc^{*}_{2(p-q)+1}\right]w_{m,2q} \\
  \nonumber && +m\pi \sum_{l=1}^{m}K_l\sum_{q=0}^{p} w_{m-l,2q} \mathcal{A}_{l,2(p-q)}+m\pi \sum_{l=m+1}^{\infty}K_l\sum_{q=0}^{m+p-l} w^{*}_{l-m,2q} \mathcal{A}_{l,2(m+p-l-q)}\\
   && \left.-m\pi \sum_{l=1}^{\infty}K_l\sum_{q=0}^{p-l} w_{m+l,2q} \mathcal{A}_{l,2(p-l-q)}  + \frac{m K_m}{2} g(\omega) \mathcal{A}_{m,2p}\right], \label{eq: recursion general}
\end{eqnarray}
and $w_{0,2p} = 0~\forall~p$ and $w_{1,0} = \psi_1(\omega)$. With these notations, Eqs.~\eqref{eq: SUP C3 EXP} and \eqref{eq: SUP C5 EXP} become
\begin{eqnarray}
    \hspace{-0.6cm}c_3 = && \pi\left[K_1\mathcal{B}_{2,0}-K_2 \mathcal{C}^{*}_{1,0}\mathcal{A}_{2,0}\right], \label{eq: SUP C3 EXP F}\\
    \hspace{-0.6cm}c_5 = && \pi
    \left[K_1 \left( \mathcal{B}_{2,2}+\mathcal{A}_{1,2}^{*}\mathcal{B}_{2,0} \right)
    -K_2 \left(\mathcal{C}^{*}_{1,0}\mathcal{A}_{2,2}+\mathcal{C}^{*}_{1,2}\mathcal{A}_{2,0}-\mathcal{A}^{*}_{2,0}B_{3,0}\right)-K_3 \mathcal{C}^{*}_{2,0}\mathcal{A}_{3,0}\right].\label{eq: SUP C5 EXP F}
\end{eqnarray}
Equation~\eqref{eq:reduced sup} is Eq.~(5) of the main text.

\section{Derivation of Eq.~(6) of the main text}

Let us start from the reduced equation~\eqref{eq:reduced sup} and define $  A \equiv r e^{i\psi} = x+iy$. Hence, we have $r=|A|$. We now compute the steady-state distribution of $r$. Decomposing Eq.~\eqref{eq:reduced sup} into its real and imaginary components, we obtain
\begin{eqnarray}
    \frac{dx}{dt} &=& \lambda_1 x -\sum_{n=1}^{\infty}c_{2n+1} x\left(x^2+y^2\right)^{n} +  \sqrt{\frac{D_\mathrm{eff}}{4\pi^2 N}} \xi_\mathrm{R} (t), \label{eq: reduced sup real} \\
    \frac{dy}{dt} &=& \lambda_1 y -\sum_{n=1}^{\infty}c_{2n+1} y\left(x^2+y^2\right)^{n} +  \sqrt{\frac{D_\mathrm{eff}}{4\pi^2 N}} \xi_\mathrm{I} (t), \label{eq: reduced sup imaginary}
\end{eqnarray}
with $\xi(t)=\frac{1}{\sqrt{2}}\xi_\mathrm{R} (t)+i\frac{1}{\sqrt{2}}\xi_\mathrm{I} (t)$ with $\langle\xi_\mathrm{R} (t)\xi_\mathrm{R} (t') \rangle=\langle\xi_\mathrm{I} (t)\xi_\mathrm{I} (t') \rangle=\delta(t-t')$ and $\langle\xi_\mathrm{R} (t) \rangle=\langle\xi_\mathrm{I} (t) \rangle=\langle\xi_\mathrm{I} (t)\xi_\mathrm{R} (t') \rangle=0$. 
The drift term in Eqs.~\eqref{eq: reduced sup real}~and~\eqref{eq: reduced sup imaginary} are negative of the gradient of the function
\begin{equation}
V(x,y) =\sum_{n\geq 0} \frac{c_{2n+1}}{2(n+1)}(x^2+y^2)^{n+1},
\end{equation}
where we have defined $c_1\equiv-\lambda_1$ to shorten the equations.
Hence, the steady state probability distribution for \eqref{eq: reduced sup real}- \eqref{eq: reduced sup imaginary} is
\begin{equation}
P(x,y) = \mathcal{N} {\rm exp}\left(N\frac{8\pi^2}{D_{\rm eff}} V(x,y)\right), 
\end{equation}
where $\mathcal{N}$ is a normalization. Performing the polar change of variable from $(x,y)$ to $(r,\theta)$, and noting that $P$ is rotationally symmetric, we obtain the steady state distribution for $r$
\begin{eqnarray}
    \mathbf{P}(r) = \mathcal{M} r \exp\Bigg\{-\frac{8\pi^2N}{D_{\rm eff}}\sum_{n=0}^{\infty}\frac{c_{2n+1}r^{2(n+1)}}{2(n+1)}\Bigg\}, \label{eq: Prob r}
\end{eqnarray}
where $\mathcal{M}$ is a normalization constant. The above is Eq.~(6) of the main text.

\section{Results corresponding to the model in \textit{Application 1} of the main text \label{sec: sup 3}}

In \textit{Application 1: Stochastic Kuramoto model with harmonic
and bi-harmonic interaction and without frequency}, we consider
\begin{eqnarray}
    K_m = 0~\forall~m\geq3,~~\mathrm{and}~~g(\omega) = \delta(\omega).
\end{eqnarray}
In this case, we could remove altogether the variable $\omega$, working with functions of $\theta$ only. However, in order to fit all applications in the same framework and apply the general formulas, we shall keep the variable $\omega$; in any case, the computations simplify a lot. 
Following Eq.~\eqref{eq: Spectral L}, the spectral function becomes
\begin{eqnarray}
    \Lambda_m(x) = 1-\frac{|m| K_{|m|}}{2\left(x+m^2D\right)},~~\mathrm{with}~~|m|=1,2.
\end{eqnarray}
The roots of this spectral function give the eigenvalues, which read as
\begin{eqnarray}
    \lambda_{\pm1} = \frac{K_1}{2}-D,~~~~~\lambda_{\pm2} = K_2-4D.~~~~~
     \label{eq: SUP Eig no freq}
\end{eqnarray}
For $|m|>2$ we obtain $\lambda_{m}=-m^2D$. The eigenfunctions are the complex exponentials: $\Psi_m =\psi_1(\omega)e^{im\theta} = \delta(\omega) e^{im\theta},~m\in \mathbb{Z}$. 

We are interested in the transition of $R_1$. Hence, the relevant eigenfunction is $\Psi_{\pm1}$. Corresponding adjoint eigenfunctions, properly normalized, are:
\begin{eqnarray}
    \tilde{\Psi}_{\pm1} = \frac{1}{2\pi}\tilde{\psi}_{\pm1}(\omega)e^{\pm i\theta} = \frac{1}{2\pi} \left(\frac{K_1}{K_1 \mp i2\omega}\right)e^{\pm i\theta}. 
\end{eqnarray}
 Since $\tilde{\Psi}_{\pm1}$ will always appear in conjunction with a $\delta(\omega)$ factor, its apparent dependency on the variable $\omega$ will have no influence on the following computations.
Evaluating the relevant expressions in Eq.~\eqref{eq: SUP C3 EXP F}, we obtain
\begin{eqnarray}
\mathcal{C}_{1,0} = 1,~\mathcal{A}_{2,0} = \mathcal{B}_{2,0} = \frac{2\pi K_1}{K_1-K_2+2D}.
\end{eqnarray}
Using these expressions, we obtain $c_3$ from Eq.~\eqref{eq: SUP C3 EXP F} as
\begin{align}
    c_3 = 2\pi^2 K_1 \left( \frac{K_1 -  K_2 }{K_1-K_2+2D}\right). \label{eq: C3 App 1 SUP}
\end{align}
Using the expression of $c_3$, we further obtain
\begin{eqnarray}
    \mathcal{A}_{1,2} &=& 0,~~  \mathcal{A}_{2,2} = \mathcal{B}_{2,2} =  \frac{8\pi^3 K_1^2 \left[    K_2^2 -K_2\left(K_1+6D\right)+2DK_1 \right]}{(K_1+4D)(K_1-K_2+2D)^2(2K_1-K_2)}, \nonumber \\
    \mathcal{C}_{1,2} &=& 0,~~\mathcal{B}_{3,0} = \frac{4\pi^2 K_1 \left(K_1+K_2\right)}{\left(K_1+4D\right)\left(K_1-K_2+2D\right)}.
\end{eqnarray}
Using all of these in Eq.~\eqref{eq: SUP C5 EXP F}, we obtain
\begin{eqnarray}
    c_5 = \frac{8\pi^4K_1^2\left[2D\left(K_1-3K_2\right)\left(K_1-K_2\right)+K_2\left(K_1^2+3K_1K_2-2K_2^2\right)\right]}{(K_1+4D)(K_1-K_2+2D)^2(2K_1-K_2)}. \label{eq: SUP C5 No Freq}
\end{eqnarray}
From the definition of $D_\mathrm{eff}$, we further get
\begin{eqnarray}
    D_\mathrm{eff} = D\int_{-\infty}^{+\infty} d\omega'\left|\tilde{\psi}_{1}(\omega')\right|^2g(\omega')=  D\int_{-\infty}^{+\infty} d\omega'\left|\frac{K_1}{K_1 \mp i2\omega'}\right|^2\delta(\omega') = D.
\end{eqnarray}

For this particular model, we may simplify the recursion relation, Eq.~\eqref{eq: recursion general}. Since we have $K_{|m|} = 0~\forall~|m|>2$, we may simplify Eq.~\eqref{eq: recursion general} for $m=1$ as
\begin{eqnarray}
    \nonumber w_{1,2p} =&&\left.\left[ \frac{1}{(1+2p)\lambda_1+D+i\omega}\right] \right[  \sum_{q=0}^{p-1}\left[(1+q)c_{2(p-q)+1}+qc^{*}_{2(p-q)+1}\right]w_{1,2q} \\
  \nonumber &&\left. +\pi K_2\sum_{q=0}^{p-1} w^{*}_{1,2q} \mathcal{A}_{2,2(m+p-l-q)}-\pi \sum_{l=1}^{2}K_l\sum_{q=0}^{p-l} w_{1+l,2q} \mathcal{A}_{l,2(p-l-q)}  + \frac{K_1}{2} g(\omega) \mathcal{A}_{1,2p}\right], \label{eq: w(1,2p) model 1}
\end{eqnarray}
and for $m>2$ as 
\begin{eqnarray}
    \nonumber w_{m,2p} =&&\left.\left[ \frac{1}{(m+2p)\lambda_1+m^2D+im\omega}\right] \right[  \sum_{q=0}^{p-1}\left[(m+q)c_{2(p-q)+1}+qc^{*}_{2(p-q)+1}\right]w_{m,2q} \\
  \nonumber && +m\pi \sum_{l=1}^{2}K_l\sum_{q=0}^{p} w_{m-l,2q} \mathcal{A}_{l,2(p-q)} \left.-m\pi \sum_{l=1}^{2}K_l\sum_{q=0}^{p-l} w_{m+l,2q} \mathcal{A}_{l,2(p-l-q)}  + \frac{m K_m}{2} \delta(\omega) \mathcal{A}_{m,2p}\right], \label{eq: w(m,2p) model 1}
\end{eqnarray}
where we have used the result $w_{0,2p} = 0~\forall~p$ and $\lambda_1 = \lambda_1^{*}$ for this model. Now we have $w_{0,2p} =0~\forall~p$ and $w_{1,0} = \psi_1(\omega) = \delta(\omega)$. This immediately gives $\mathcal{A}_{1,0} =\mathcal{B}_{1,0}=\mathcal{C}_{1,0}= 1$. We now focus on obtaining $w_{m,0}$. Putting $p=0$ in Eq.~\eqref{eq: w(m,2p) model 1}, we obtain for $m\geq 2$ that
\begin{equation}
    w_{m,0} =\Bigg[ \frac{1}{m\lambda_1+m^2D+im\omega}\Bigg] \left[m\pi K_1 w_{m-1,0} \mathcal{A}_{1,0}+m\pi K_2 w_{m-2,0} \mathcal{A}_{2,0}+ \frac{m K_m}{2} \delta(\omega) \mathcal{A}_{m,0}\right] .\label{eq: w(m,0) model 1}
\end{equation}
Hence, for $m=2$, we have
\begin{equation}
    w_{2,0} =\Bigg[ \frac{1}{2\lambda_1+4D+i2\omega}\Bigg] \left[2\pi K_1 w_{1,0} \mathcal{A}_{1,0}+K_2 \delta(\omega) \mathcal{A}_{2,0}\right]   = \Bigg[ \frac{2\pi K_1 \mathcal{A}_{1,0}+K_2  \mathcal{A}_{2,0}}{2\lambda_1+4D}\Bigg]\delta(\omega).\label{eq: w(2,0) model 1}
\end{equation}
Integrating both sides with respect to $\omega$ and noting that $\mathcal{A}_{2,0} = \int_{-\infty}^{\infty}d\omega w_{2,0}$, we obtain
\begin{eqnarray}
     w_{2,0} = \mathcal{A}_{2,0}\delta(\omega) = \left[\frac{2\pi K_1}{K_1-K_2+2D}\right]\delta(\omega).
\end{eqnarray}
Since $w_{1,0}$ and $w_{2,0}$ are proportional to $\delta(\omega)$, then using Eq.~\eqref{eq: w(m,0) model 1} we deduce that any $w_{m,0}$ is proportional to $\delta(\omega)$. Hence, we may write $w_{m,0} = \mathcal{A}_{m,0}\delta(\omega)$ and use this in Eq.~\eqref{eq: w(m,0) model 1} to obtain
\begin{eqnarray}
    w_{m,0} =\pi\Bigg[ \frac{ K_1 \mathcal{A}_{m-1,0} + K_2 \mathcal{A}_{m-2,0} \mathcal{A}_{2,0}}{\lambda_1+mD}\Bigg] \delta(\omega),
\end{eqnarray}
where $\mathcal{A}_{m,0}$'s can be obtained from the recursion relation as
\begin{eqnarray}
    \mathcal{A}_{m,0} = \pi\Bigg[ \frac{ K_1 \mathcal{A}_{m-1,0} + K_2 \mathcal{A}_{m-2,0} \mathcal{A}_{2,0}}{\lambda_1+mD}\Bigg],
\end{eqnarray}
for $m\geq2$, and $\mathcal{A}_{2,0}$ and $\mathcal{A}_{1,0}$ are known.

Let us now focus on Eq.~\eqref{eq: w(m,2p) model 1}. We observe that any $w_{m,2p}$ is expressible as a linear combination of those $w_{m',2p'}$ that satisfy $m'+2p'\leq m+2p$ with $m' = m-2,m-1,m,m+1$, and $m+1$. Now, if all those $w_{m',2p'}$ have a form $w_{m',2p'} = \mathcal{A}_{m',2p'}\delta(\omega)$, where $\mathcal{A}_{m',2p'} = \int_{-\infty}^{\infty}d\omega w_{m',2p'}$ is a constant, then $w_{m,2p}$ can also be written as
\begin{eqnarray}
    w_{m,2p} = \mathcal{A}_{m,2p} \delta(\omega),\label{eq: w(m,2p) result one}
\end{eqnarray}
with the recursion relation of $\mathcal{A}_{m,2p}$'s reading for $m=1$ as
\begin{eqnarray}
 \mathcal{A}_{1,2p} &=& \frac{1}{2p\lambda_1}\Bigg[  \sum_{q=0}^{p-1}(1+2q)c_{2(p-q)+1}\mathcal{A}_{1,2q} +\pi K_2\sum_{q=0}^{p-1} \mathcal{A}_{1,2q} \mathcal{A}_{2,2(p-q-1)}-\pi K_1\sum_{q=0}^{p-1} \mathcal{A}_{2,2q} \mathcal{A}_{1,2(p-q-1)} \nonumber\\
 &&-\pi K_2\sum_{q=0}^{p-2} \mathcal{A}_{3,2q} \mathcal{A}_{2,2(p-q-2)} \Theta(p-2)\Bigg], \label{eq: A(1,2p) model 1}
\end{eqnarray}
and for $m>2$ as
\begin{eqnarray}
    \nonumber \mathcal{A}_{m,2p} =&&\left.\left[ \frac{1}{(m+2p)\lambda_1+m^2D}\right] \right[  \sum_{q=0}^{p-1}\left[(m+2q)c_{2(p-q)+1}\right]\mathcal{A}_{m,2q} \\
   &&\left. +m\pi \sum_{l=1}^{2}K_l\sum_{q=0}^{p} \mathcal{A}_{m-l,2q} \mathcal{A}_{l,2(p-q)} -m\pi \sum_{l=1}^{2}K_l\sum_{q=0}^{p-l} \mathcal{A}_{m+l,2q} \mathcal{A}_{l,2(p-l-q)}  + \frac{m K_m}{2}  \mathcal{A}_{m,2p}\right]. \label{eq: A(m,2p) model 1}
\end{eqnarray}

From Eq.~\eqref{eq: w(m,2p) result one}, we immediately have $\mathcal{A}_{m,2p}=\mathcal{B}_{m,2p}=\mathcal{C}_{m,2p} \in \mathbb{R}~\forall~m,p$ for this model. Hence, we can simplify Eq.~\eqref{eq: c 2n+1 general} for this model as
\begin{equation}
   c_{2n+1} = \pi  K_1\sum_{q=0}^{n-1}\mathcal{A}_{2,2q}\mathcal{A}_{1,2(n-q-1)} +\pi K_2\sum_{q=0}^{n-2}\mathcal{A}_{3,2q}\mathcal{A}_{2,2(n-q-2)} \Theta(n-q-2)-\pi K_2\sum_{q=0}^{n-1}\mathcal{A}_{1,2q}\mathcal{A}_{2,2(n-q-1)}. \label{eq: c 2n+1 model 1}
\end{equation}

Let us now focus on $\mathcal{A}_{1,2p}$'s. We already have $\mathcal{A}_{1,0} = 1$. Now, from Eq.~\eqref{eq: A(1,2p) model 1}, we obtain
\begin{equation}
 \mathcal{A}_{1,2} = \frac{1}{2p\lambda_1}\left[  c_{3} +\pi K_2 \mathcal{A}_{1,0} \mathcal{A}_{2,0}-\pi K_1 \mathcal{A}_{2,0} \right] = 0 , \label{eq: A(1,2) model 1}
\end{equation}
since by definition $c_3 = \pi\left[K_1\mathcal{B}_{2,0}-K_2 \mathcal{C}^{*}_{1,0}\mathcal{A}_{2,0}\right] =  \pi\left[K_1\mathcal{A}_{2,0}-K_2 \mathcal{A}_{1,0}\mathcal{A}_{2,0}\right]$. Let us now prove by induction that $\mathcal{A}_{1,2p} = 0~\forall~p>0$. We have already proved that $\mathcal{A}_{1,2} = 0$. Now, let us assume that $\mathcal{A}_{1,4} = \mathcal{A}_{1,6} = \ldots=\mathcal{A}_{1,2(p-1)} = 0$. Putting this condition in Eq.~\eqref{eq: A(1,2p) model 1}, we obtain
\begin{eqnarray}
 \mathcal{A}_{1,2p} &=& \frac{1}{2p\lambda_1}\Bigg[  c_{2p+1}-\pi K_1 \mathcal{A}_{2,2(p-1)} -\pi K_2\sum_{q=0}^{p-2} \mathcal{A}_{3,2q} \mathcal{A}_{2,2(p-q-2)} \Bigg] = 0.
\end{eqnarray}
In the last step, we have used the expression of $c_{2p+1}$ from Eq.~\eqref{eq: c 2n+1 model 1} with the condition $\mathcal{A}_{1,2m}=0~\forall~m\in \{1,2,\ldots,p-1\}$. 
Hence, by the method of induction, we have proved $\mathcal{A}_{1,2p} = 0~\forall~p\geq1$.

Using these results, we may simplify Eqs.~\eqref{eq: c 2n+1 general}~and~\eqref{eq: recursion general} as
\begin{eqnarray}
    c_{2n+1} = \pi \left[ (K_1-K_2) \mathcal{A}_{2,2(n-1)} +K_2 \sum_{j=0}^{n-2}\mathcal{A}_{2,2j}~ \mathcal{A}_{3,2(n-j-2)}\right],
\end{eqnarray}
where
\begin{eqnarray}
    \nonumber w_{m,2l} = \frac{1}{\left[(m+2l)\lambda_1+m^2D\right]} &&\left[\sum_{j=0}^{l-1}(m+2j)c_{2(l-j)+1}~ \mathcal{A}_{m,2j}+m\pi K_1 \left(\mathcal{A}_{m-1,2l}-\mathcal{A}_{m+1,2(l-1)}\right)\right.\\
    &&~~~~~~~\left. +m\pi K_2 \sum_{j=0}^{l} \mathcal{A}_{2,2j} \left(\mathcal{A}_{m-2,2(l-j)}-\mathcal{A}_{m+2,2(n-j-2)}\Theta(n-j-2)\right)\right],
\end{eqnarray}
for $n = 1, 2,3, \ldots$ and with $w_{0,2l} = 0~\forall~l$ and $w_{1,0} = 1$ and $w_{1,2l} = 0~\forall~l>0$ and $\Theta(x) = 1~\forall~x\geq0$ and $\Theta(x) = 0~\forall~x<0$. Using these relations recursively, we may calculate $ c_{2n+1}$ for any $n$.

\section{Results corresponding to the model in \textit{Application 2} of the main text \label{sec: sup 4}}

In \textit{Application 2: Stochastic Kuramoto model with harmonic
and bi-harmonic interaction with bi-delta frequency}, we consider
\begin{eqnarray}
    K_m = 0~\forall~m\geq3,~~\mathrm{and}~~g(\omega) = \frac{1}{2}\delta(\omega-\omega_0)+\frac{1}{2}\delta(\omega+\omega_0).
\end{eqnarray}
As discussed in the main text, when $N$ is finite, the density $\bar{g}_{_N}(\omega) = \sum_{j=1}^N\delta(\omega-\omega_j)$ will not exactly match with $g(\omega)$. However, we may in any case write $\bar{g}_{_N}(\omega) = \alpha\delta(\omega-\omega_0)+(1-\alpha)\delta(\omega+\omega_0)$, incorporating in this case the finite-size effects from frequency disorder.
For this specific frequency distribution, we could simplify the dependency on the $\omega$ variable by using two-components distribution functions, for $\pm \omega_0$. However, as for application 1, in order to apply the general formulas, we shall keep the variable $\omega$ as if it were continuous. In a typical frequency realization, the finite-size parameter $\alpha$ is such that $\alpha\to 1/2$ as $N\to \infty$. Following Eq.~\eqref{eq: Spectral L}, the spectral function becomes
\begin{eqnarray}
    \Lambda_m(x) = 1-\frac{|m| K_{|m|} \left[ x+m^2D+i(1-2\alpha)m\omega_0\right]}{2\left[\left(x+m^2D\right)^2+m^2\omega_0^2\right]},~~\mathrm{with}~~|m|=1,2.
\end{eqnarray}
The roots of this spectral function give the eigenvalues, which read as
\begin{align}
    \lambda_{m,\pm} &= -m^2D +
   \frac{|m| K_m}{2}\left[\frac{1}{2}\pm \sqrt{\left(\frac{1}{4} - \frac{4\omega_0^2}{K_m^2}\right)+i2(1-2\alpha) \frac{|m|\omega_0}{mK_m}}\right],~~\mathrm{with}~~|m|=1,2. \label{eq: eig model 1 sup}
\end{align}
The eigenfunctions corresponding to these eigenvalues are given by Eq.~\eqref{eq: Psim sup}. Putting $\omega_0=0$, we get back the eigenvalues given in Eq.~\eqref{eq: SUP Eig no freq}. For $|m|\geq3$, the eigenvalues are $\lambda_{m,\pm} = -m^2D\pm im\omega_0$ with the corresponding eigenfunction $\psi_{m,\pm}(\omega) = \delta(\omega \pm \omega_0)$. 
The spectral analysis for $|m|\geq 3$ will not be needed.
 Note that, unlike the previous example, here for each Fourier mode $m=\pm 1,\pm 2$, there are two eigenvalues and two corresponding  eigenfunctions, denoted by $\pm$. In this case the two eigenfunction of $\mathcal{L}$ corresponding to the $m$th mode may be written as, following Eq.~\eqref{eq: Psim sup}
\begin{eqnarray}
    \psi_{m,\pm}(\omega) = \frac{|m| K_{|m|} }{4\left(\lambda_{m,\pm}+m^2D+im\omega_0\right)} \delta(\omega-\omega_0) +\frac{|m| K_{|m|} }{4\left(\lambda_{m,\pm}+m^2D-im\omega_0\right)} \delta(\omega+\omega_0). 
\end{eqnarray}
Clearly, the two eigenfunctions $\psi_{m,\pm}(\omega)$ lie on the two-dimensional space spanned by $\delta(\omega-\omega_0)$ and $\delta(\omega+\omega_0)$. Following Eq.~\eqref{eq: SUP psi tilde}, we have the eigenfunctions of $\mathcal{L}^\dagger$ which read
\begin{eqnarray}
    \tilde{\psi}_{1,\pm}(\omega) =\frac{\mathcal{G}^{*}_\pm}{(\lambda^{*}_{1,\pm}+D-i\omega)} ,~~\mathcal{G}_{\pm}  = \frac{2\left[\left(\lambda_{1,\pm}+D\right)^2+\omega_0^2\right]^2}{K_{1} \left[\left(\lambda_{1,\pm}+D\right)^2-\omega_0^2+ i 2(1-2\alpha)\omega_0 \left(\lambda_{1,\pm}+D\right)\right]}.
\end{eqnarray}

As discussed in the main text, in the region of our interest, the relevant unstable mode is $\psi_{1,+}$. In this case, evaluating the relevant expressions in Eq.~\eqref{eq: SUP C3 EXP F}, we obtain
\begin{eqnarray}
    \mathcal{C}^{*}_{1,0} &=&\frac{K_1}{2} \mathcal{G}_+\mathcal{ H},~~\mathrm{where}~~\mathcal{H}\equiv \frac{\left( \lambda_{1,+} + D\right)\left( \lambda^{*}_{1,+}  + D \right)+\omega_0^2+ i (1-2\alpha) \omega_0 \left(\lambda^{*}_{1,+}-\lambda_{1,+}\right)}{\left|\left[\left( \lambda_{1,+} + D\right)^2+\omega_0^2 \right]\right|^2},\\
    \mathcal{A}_{2,0} &=& \frac{\pi K_1^2\left[\left(\lambda_{1,+}+2D \right)\left(\lambda_{1,+}+D \right)-\omega_0^2+ i (1-2\alpha)\omega_0 \left(2 \lambda_{1,+}+3D \right)\right]}{2\left[\left(\lambda_{1,+}+D \right)^2+\omega_0^2\right]\left[\left(\lambda_{1,+}+2D \right)^2+\omega_0^2  - \left.\left.\left(\frac{K_2}{2}\right) \right(\lambda_{1,+}+2D + i(1-2\alpha)\omega_0 \right)\right]},\\
    \mathcal{B}_{2,0}&=& \frac{\pi K_1^2}{2} \mathcal{G}_+\mathcal{ I} + \frac{K_2}{2}\mathcal{G}_+\mathcal{A}_{2,0}\mathcal{K}~~\mathrm{where},~~\mathcal{K} \equiv \frac{\left(\lambda_{1,+}+2D \right)\left(\lambda_{1,+}+D \right)-\omega_0^2+ i (1-2\alpha)\omega_0 \left(2 \lambda_{1,+}+3D \right)}{\left[\left(\lambda_{1,+}+2D \right)^2+\omega_0^2 \right]\left[\left(\lambda_{1,+}+D \right)^2+\omega_0^2\right]},
\end{eqnarray}
and
\begin{eqnarray}
    \hspace{-1cm}\mathcal{I}&\equiv& \frac{\left[\left( \lambda_{1,+}  + D \right)^2\left(\lambda_{1,+}+2D \right) -\omega_0^2 \left(3\lambda_{1,+}+4D \right)\right]+i (1-2\alpha)\omega_0\left[\left( \lambda_{1,+}  + D \right)^2 -\omega_0^2+2\left( \lambda_{1,+}  + D \right)\left(\lambda_{1,+}+2D \right)\right]}{\left[\left( \lambda_{1,+}  + D\right)^2+\omega^2_0\right]^2\left[\left(\lambda_{1,+}+2D \right)^2+\omega_0^2\right]},
\end{eqnarray}
Using these expressions, we obtain $c_3$ from Eq.~\eqref{eq: SUP C3 EXP F}, which reads as
\begin{eqnarray}
    c_3=\frac{\pi K_1}{2}\mathcal{G}_+\left[\pi K_1^2  \mathcal{I} + K_2  \mathcal{A}_{2,0} \left( \mathcal{K}-\mathcal{H}\right)\right]. \label{eq: C3 App 2 SUP}
\end{eqnarray}
Putting $\omega_0 = 0$ in Eq.~\eqref{eq: C3 App 2 SUP}, we get back Eq.~\eqref{eq: C3 App 1 SUP}. To write the expression of $c_5$, let us define a functional $\mathbb{P}[x(\cdot)] = \frac{\alpha}{x(\omega_0)} +\frac{1-\alpha}{x(-\omega_0)}$, to make the expressions look neat. Here, given that $x(\omega_0)$ if a function of $\omega_0$, the quantity $x(-\omega_0)$ is what we obtain replacing $\omega_0$ by $-\omega_0$ in the expression of $x(\omega_0)$. In this way, we may write
\begin{eqnarray}
    &&\mathcal{A}_{1,2} = \frac{\mathcal{D}_1}{1-\mathcal{D}_0},~\mathrm{where}~\mathcal{D}_0= \frac{K_1}{2} \mathbb{P}\left[x_5\right],~\mathcal{D}_1 =\frac{K_1}{2}c_3\mathbb{P}\left[x_1x_5\right]-\frac{\pi^2K^3_1}{2}\mathbb{P}\left[x_1x_2x_5\right] -\frac{\pi K_1 K_2}{2}\mathcal{B} \left\{\mathbb{P}\left[x_2x_5\right]-\mathbb{P}\left[x_{-1}x_5\right]\right\},\nonumber \\ \\
    &&\mathcal{A}_{2,2} = \frac{\mathcal{E}_1}{1-\mathcal{E}_0},~~\mathrm{where}~~\mathcal{E}_0= K_2 \mathbb{P}\left[x_4\right],\\
    &&\mathcal{E}_1= \pi K_1^2\mathcal{A}_{1,2} \left \{ \mathbb{P}\left[x_1x_4\right] +\mathbb{P}\left[x_4x_5\right]\right\}+ \pi K_1^2c_3 \left\{\mathbb{P}\left[x_1x_2x_4\right]+\mathbb{P}\left[x_1x_4x_5\right]\right\}-\pi^3K_1^4 \left\{\mathbb{P}\left[x_1x_2x_3x_4\right] +\mathbb{P}\left[x_1x_2x_4x_5\right]\right\}\nonumber\\
    && +K_2 c_3 \mathcal{A}_{2,0} \mathbb{P}\left[x_2x_4\right] -\pi^2K_1^2K_2 \mathcal{A}_{2,0}\left\{\mathbb{P}\left[x_1x_3x_4\right] +\mathbb{P}\left[x_2x_3x_4\right]+\mathbb{P}\left[x_2x_4x_5\right]-\mathbb{P}\left[x^{*}_1x_4x_5\right]\right\},
\end{eqnarray}
where $x_1 (\omega_0)= \left(\lambda_{1,+}+D+i\omega_0\right),~x_{-1} (\omega_0)= \left(\lambda_{1,+}+D-i\omega_0\right),~x_2(\omega_0) = \left(\lambda_{1,+}+2D+i\omega_0\right),~x_3(\omega_0) = \left(\lambda_{1,+}+3D+i\omega_0\right)$, and moreover that $x_4(\omega_0) = \left(3\lambda_{1,+}+\lambda_{1,+}^{*}+4D+i2\omega_0\right)$ and $x_5(\omega_0) = \left(2\lambda_{1,+}+\lambda_{1,+}^{*}+D+i\omega_0\right)$. In their terms, we may express the rest of the integrals, which read as
\begin{eqnarray}
    \mathcal{B}_{2,2} =&& \pi K_1^2 \mathcal{A}_{1,2}\mathcal{G}_+ \left\{\mathbb{P}\left[x_1x_4\right]+\mathbb{P}\left[x_1x_4x_5\right]\right\}+\pi K_1^2c_3\mathcal{G}_+ \left\{\mathbb{P}\left[x^2_1x_2x_4\right]+\mathbb{P}\left[x^2_1x_4x_5\right]\right\}-\pi^3K_1^4\mathcal{G}_+ \left\{\mathbb{P}\left[x^2_1x_2x_3x_4\right]\right. \nonumber \\
    &&\left.+\mathbb{P}\left[x^2_1x_2x_4x_5\right]\right\}+K_2\mathcal{A}_{2,2}\mathcal{G}_+ \mathbb{P}\left[x_1x_4\right]+K_2c_3\mathcal{A}_{2,0}\mathcal{G}_+ \mathbb{P}\left[x_1x_2x_4\right] -\pi^2K_1^2K_2\mathcal{A}_{2,0}\mathcal{G}_+ \left\{ \mathbb{P}\left[x_1^2x_3x_4\right]+ \mathbb{P}\left[x_1x_2x_3x_4\right]  \right. \nonumber\\
    &&\left. +\mathbb{P}\left[x_1x_2x_4x_5\right]-\mathbb{P}\left[x_1x^{*}_1x_4x_5\right] \right\},\\
    \mathcal{C}^{*}_{1,2} =&& \frac{K_1}{2} c_3 \mathcal{G}_+^{*} \mathbb{P}\left[x_1x_1^{*}x_5\right] -\frac{\pi^2 K^3_1}{2}  \mathcal{G}_+^{*} \mathbb{P}\left[x_1x_1^{*}x_2x_5\right] +\frac{\pi^2 K_1K_2}{2}  \mathcal{A}_{2,0}\mathcal{G}_+^{*} \left\{\mathbb{P}\left[\left(x_1^{*}\right)^2x_5\right]+\mathbb{P}\left[x_1^{*}x_2x_5\right]\right\}\nonumber\\
    &&+\frac{K_1}{2}\mathcal{A}_{1,2}\mathcal{G}_+^{*}\mathbb{P}\left[x^{*}_1x_5\right],\\
    \mathcal{B}_{3,0} = && \frac{\pi^2 K_1^3}{2}\mathcal{G}_+\mathbb{P}\left[x_1^2x_2x_3\right] +\frac{\pi K_1 K_2}{2} \mathcal{A}_{2,0}\mathcal{G}_+ \left\{\mathbb{P}\left[x_1^2x_3\right]+\mathbb{P}\left[x_1x_2x_3\right] \right\}.
\end{eqnarray}
Putting all of these together into Eq.~\eqref{eq: SUP C5 EXP F}, we obtain $c_5$. Putting $\omega_0 =0$ into the final expression, we get back Eq.~\eqref{eq: SUP C5 No Freq}. From the definition of $D_\mathrm{eff}$, we further get
\begin{eqnarray}
    D_\mathrm{eff} = D \int_{-\infty}^{+\infty} d\omega'\left|\tilde{\psi}_{1}(\omega')\right|^2g(\omega')= D \int_{-\infty}^{+\infty} d\omega'\left|\frac{\mathcal{G}^{*}_+}{(\lambda^{*}_{1,+}+D-i\omega')}\right|^2g(\omega')=D \left|\mathcal{G}_+\right|^2 \mathcal{H}. \label{eq: D effective sup}
\end{eqnarray}

For $\alpha = 1/2$, the case considered in the main text, many of these expressions simplify. Firstly, the eigenvalues of the operator $\mathcal{L}$, given by Eq.~\eqref{eq: eig model 1 sup}, simplify to
\begin{align}
    \lambda_{1,\pm} &= -D +
   \frac{ K_1}{4}\pm \sqrt{\left(\frac{K_1}{4}\right)^2 -\omega_0^2 }.
\end{align}
Now we are interested in determining the nature of transition at the transition point. In other words, we need to determine the sign of $c_3$ at the transition point. At the transition point, we have $\lambda_{1,+} = 0$. Putting $\lambda_{1,+} =0$ and $\alpha = 1/2$ into previously obtained expressions, we obtain
\begin{eqnarray}
    &&\mathcal{G}_+ = \frac{2(D^2+\omega_0^2)^2}{K_1(D^2-\omega_0^2)},~~\mathcal{I} = \frac{2D^3-4\omega^2_0D}{\left(D^2+\omega_0^2\right)^2\left( 4D^2+\omega_0^2\right)},~~\mathcal{A}_{2,0} = \frac{ K_1^2}{2} \frac{2D^2-\omega_0^2}{\left[D^2+\omega_0^2\right]\left[4D^2+\omega_0^2-K_2D\right]},\nonumber \\
  &&    \mathcal{K} = \frac{2D^2-\omega_0^2}{\left[4D^2+\omega_0^2\right]\left[D^2+\omega_0^2\right]},~~\mathcal{H} = \frac{1}{D^2+\omega_0^2}.
\end{eqnarray}
Combining all of them, we obtain 
\begin{eqnarray}
    c_3  = \frac{K^2_1 D}{\left(D^2+\omega_0^2\right)^2\left( 4D^2+\omega_0^2\right)} \frac{\left(2D^2-4\omega_0^2\right)\left(4D^2+\omega_0^2\right)-DK_2\left(4D^2-5\omega_0^2\right)}{4D^2+\omega_0^2-DK_2}.
\end{eqnarray}

From the discussions and Figure 3(a) of the main text, the region of validity of our calculation with a two-dimensional unstable subspace is $\omega_0 <D$ and $K_2 <K_2^\mathrm{c}=4D+\omega_0^2/D$. Hence, we have $(4D^2+\omega_0^2-DK_2)>0$ for the entire region of validity of our calculation. As a result, the sign of $c_3$ will be determined from the sign of $\left[\left(2D^2-4\omega_0^2\right)\left(4D^2+\omega_0^2\right)-DK_2\left(4D^2-5\omega_0^2\right)\right]$. Hence, we may write
\begin{equation}
    c_3 \propto \left[\left(2D^2-4\omega_0^2\right)\left(4D^2+\omega_0^2\right)-DK_2\left(4D^2-5\omega_0^2\right)\right].
\end{equation}
The tri-critical point may be obtained from the condition $c_3=0$, which gives
\begin{equation}
   \omega_0 =  \frac{1}{2\sqrt{2}} \sqrt{5K_2D-14D^2 + D\sqrt{25 K^2_2-204 K_2D+324D^2} }. \label{eq: omeha0 k2 sup}
\end{equation}
Equation~\eqref{eq: omeha0 k2 sup} gives the black line in Figure~(3) panel (b) of the main text separating the continuous and first-order transition regions.

Putting $\alpha = 1/2$ in the expressions of $\mathcal{G}_+$ and $\mathcal{H}$ and putting them back in Eq.~\eqref{eq: D effective sup}, we obtain
\begin{eqnarray}
    D_\mathrm{eff} &=& D \left|\mathcal{G}_+\right|^2 \mathcal{H} = D\left|\frac{2\left[\left(\lambda_{1,+}+D\right)^2+\omega_0^2\right]^2}{K_{1} \left[\left(\lambda_{1,+}+D\right)^2-\omega_0^2\right]}\right|^2\frac{\left( \lambda_{1,+} + D\right)\left( \lambda^{*}_{1,+}  + D \right)+\omega_0^2}{\left|\left[\left( \lambda_{1,+} + D\right)^2+\omega_0^2 \right]\right|^2}= D \frac{4\left[\left(\lambda_{1,+}+D\right)^2+\omega_0^2\right]^3 }{\left[\left(\lambda_{1,+}+D\right)^2-\omega_0^2\right]^2}\nonumber \\
    &=& D \frac{4\left[ \frac{K^2_1}{8} + \frac{K_1}{2}\sqrt{\frac{K_1^2}{16}-  \omega_0^2 }\right]^3 }{\left[ \left(\frac{ K_1}{4}+ \sqrt{\left(\frac{K_1}{4}\right)^2 -\omega_0^2 }\right)^2-\omega_0^2\right]^2}.
\end{eqnarray}
Hence, we have
\begin{eqnarray}
    D_\mathrm{eff} \propto \left[ \left(\frac{ K_1}{4}+ \sqrt{\left(\frac{K_1}{4}\right)^2 -\omega_0^2 }\right)^2-\omega_0^2\right]^{-2},
\end{eqnarray}
which diverges on the line $K_1=4\omega_0$.

\section{Results corresponding to the model in \textit{Application 3} of the main text \label{sec: sup 5}}

In \textit{Application 3: Stochastic Kuramoto model with harmonic
and bi-harmonic interaction with Lorentzian frequency}, we consider
\begin{eqnarray}
    K_m = 0~\forall~m\geq3,~\mathrm{and}~g(\omega) = \frac{\sigma}{\pi\left(\omega^2+\sigma^2\right)}.
\end{eqnarray}
Following Eq.~\eqref{eq: Spectral L}, the spectral function becomes
\begin{eqnarray}
    \Lambda_m(x) = 1-\frac{|m| K_{|m|}}{2\left(x+m^2D\pm m\sigma\right)},~~\mathrm{if}~~\Re(x) \gtrless-m^2D,~~\mathrm{for}~~|m|=1,2.
\end{eqnarray}
The roots of this spectral function give the eigenvalues, which read as
\begin{align}
    \lambda_m &= |m |\left(K_m-2\sigma\right)/2 -m^2D, \qquad m = \pm1,\pm 2
\end{align}
for $K_m \geq 2\sigma$. For $m = \pm1,\pm 2$ with $K_m < 2\sigma$ and for $m = \pm3, \pm4,\ldots$, the eigenvalues are $\lambda_m = -m^2 D-im\omega$. Putting $\sigma=0$, we get back the eigenvalues given in Eq.~\eqref{eq: SUP Eig no freq}. Following Eq.~\eqref{eq: SUP psi tilde}, we have the eigenfunctions of $L^\dagger_m$ for $m=\pm1,\pm2$, which read
\begin{eqnarray}
    \tilde{\psi}_{m}(\omega) =\frac{\mathcal{G}^{*}}{(\lambda^{*}_{m,\pm}+m^2D-im\omega)} ,~~\mathcal{G}  = \frac{|m|K_{|m|}}{2 }.
\end{eqnarray}
Evaluating the relevant expressions in Eq.~\eqref{eq: SUP C3 EXP F}, we obtain
\begin{eqnarray}
    \mathcal{C}_{1,0} = \frac{K_1}{K_1-2\sigma},~~\mathcal{A}_{2,0} = \mathcal{B}_{2,0} = \frac{2\pi K_1}{K1-K_2+2D}.
\end{eqnarray}
Using these expressions, we obtain $c_3$ from Eq.~\eqref{eq: SUP C3 EXP F}, which reads as
\begin{eqnarray}
    c_3 = \frac{2\pi^2 K^2_1}{K_1-2\sigma} \left( \frac{K_1 -  K_2 -2\sigma}{K_1-K_2+2D}\right).
\end{eqnarray}
Using the expression of $c_3$, we further compute
\begin{eqnarray}
    \mathcal{A}_{1,2} &=& -\frac{2 \pi^2 \sigma K_1 \left[K_1^3-2K_1K_2\left(2D+\sigma\right)+4K_2\left(D+\sigma\right)\left(3D+2\sigma\right)-K_1^2\left(K_2+2\sigma\right)\right]}{\left(K_1-K_2+2D\right)\left(K_1-2\sigma\right)\left(K_1-2D-2\sigma\right)\left(K_1-3D-2\sigma\right)\left(K_1-2D-4\sigma\right)} ,~~\mathrm{if}~~\left(2\sigma+\frac{4}{3}D\right)>K_1>2\sigma, \nonumber\\
    &=&  - \frac{2 \pi^2 \sigma K_1K_2 }{\left(K_1-K_2+2D\right)\left(K_1-2\sigma\right)\left(K_1-D-2\sigma\right)} ,~~~~~~~~~~~~~~~~~~~~~~~~~~~~~~~~\mathrm{if}~~K_1>\left(2\sigma+\frac{4}{3}D\right).
\end{eqnarray}
In this model, the transition point is $K_1^\mathrm{c} = 2(D+\sigma)$. Since $K_1^c>2\sigma+4D/3$ and we are mostly interested in the finite-size fluctuations near the transition point,  here we compute the rest of the expressions for $K_1>\left(2\sigma+4D/3\right)$. They read as
\begin{eqnarray}
    \mathcal{A}_{2,2} &=& \frac{
2 K_1^2 \pi^2\mathcal{E}_1 }{3 (4 D_0 + K_1) (2 D_0 + K_1 - K_2)^2 (K_1 - 2 \sigma) (-D_0 + K_1 - 2 \sigma)(K_1 - \sigma)},~~\mathrm{where}\nonumber\\
\mathcal{E}_1&=&
6 K_1 (-D_0 + K_1) \left(2 D_0 K_1 - (6 D_0 + K_1) K_2 + K_2^2 \right) \pi+ 4 \left(6 D_0 (D_0 - 2K_1) K_1 \right. \nonumber \\
&&\left. + 2 (-7 D_0^2 + 9 D_0 K_1 + K_1^2) K_2 + (7 D_0 - 5 K_1) K_2^2\right) \pi \sigma + 
24 (2 D_0 (K_1 - K_2) + K_2^2) \pi \sigma^2
\\
   \mathcal{C}^{*}_{1,2} &=& - \frac{
K_1 \pi^2 \sigma \left[
- D_0 K_1 (K_1 - 2 K_2)(K_1 - 2 \sigma) + 
4 D_0^2 K_2 (-K_1 + \sigma) + 
(K_1 - 2 \sigma)^2 (K_1^2 + 4 K_1 K_2 - 4 K_2 \sigma)
\right]
}{
(2 D_0 + K_1 - K_2) (K_1 - 2 \sigma)^2 (-D_0 + K_1 - 2 \sigma)^2 (D_0 + K_1 - 2 \sigma)
},\\
\mathcal{B}_{2,2} &=& \frac{
  4 K_{1}^{2} \pi^{3} \mathcal{E}_2
}{
  3 (4 D_{0} + K_{1})(2 D_{0} + K_{1} - K_{2})^{2}(K_{1} - 2\sigma)^{2}(-D_{0} + K_{1} - 2\sigma)(2 K_{1} - K_{2} - 2\sigma)
},~~\mathrm{where}\nonumber \\
\mathcal{E}_2 &=& 
    6 K_{1}^{2} \big(-D_{0} + K_{1}\big)\big(2 D_{0} K_{1} - (6 D_{0} + K_{1})K_{2} + K_{2}^{2}\big)
    + \big(24 D_{0}(2 D_{0} - 3 K_{1})K_{1}^{2}
      + 2 K_{1}(-56 D_{0}^{2}+ 72 D_{0} K_{1}\nonumber
      \\
      &&  + 11 K_{1}^{2})K_{2} + (-8 D_{0}^{2} + 26 D_{0} K_{1} - 35 K_{1}^{2})K_{2}^{2}
      + (4 D_{0} + K_{1})K_{2}^{3}\big)\sigma- 6 \big(8 D_{0}^{2}(K_{1}- 2 K_{2})
      + K_{1}(3 K_{1} - 11 K_{2})K_{2} \nonumber \\
      &&
      + D_{0}(-24 K_{1}^{2} + 34 K_{1} K_{2} + 8 K_{2}^{2})\big)\sigma^{2}
    - 48 \big(2 D_{0}(K_{1} - K_{2}) + K_{2}^{2}\big)\sigma^{3},\\
  \mathcal{B}_{3,0} &=& \frac{4\pi^2 K_1 \left(K_1+K_2\right)}{\left(K_1+4D\right)\left(K_1-K_2+2D\right)}.
\end{eqnarray}
Putting all of these together into Eq.~\eqref{eq: SUP C5 EXP F}, we obtain $c_5$.

From the definition of $D_\mathrm{eff}$, we further get
\begin{eqnarray}
    D_\mathrm{eff} = D \int_{-\infty}^{+\infty} d\omega'\left|\tilde{\psi}_{1}(\omega')\right|^2g(\omega') = \frac{DK_1}{K_1-2\sigma}.
\end{eqnarray}
\section{Derivation of Eq. (13)}
For the model in \textit{Application 1}, we have $g(\omega) = \delta(\omega), ~\lambda_1 = K_1/2-D$ and
\begin{eqnarray}
    \Lambda'_1(x) = \frac{ K_{1}}{2\left(x+D\right)^2},
\end{eqnarray}
which immediately gives $\Lambda'_1(\lambda_1) = 2/K_1$. Hence, we further get $\psi_1 = \delta(\omega)$ and
\begin{eqnarray}
    \tilde{\psi}_1(\omega) = \frac{1}{\left[\Lambda'_1 (\lambda_{1})\right]^{*} }\frac{1}{(\lambda_{1}^{*}+D-i\omega)}  = \frac{K_1}{K_1-i2\omega}.
\end{eqnarray}
Another relevant quantity for this calculation is
\begin{equation}
    w_{2,0}(\omega)=\frac{\pi K_1 \psi_1(\omega)   }{\left(\lambda_1 + 2D+i\omega\right)} = \frac{2\pi K_1 \delta(\omega)   }{\left(K_1 + 2D\right)}. 
\end{equation}

Let us now start from Eq.~\eqref{eq: SUP ddt A detailed}. We have
\begin{eqnarray}
    \dot{A}(t) = \left( \tilde{\Psi}_1 ,\mathcal{L}\eta \right)+\left( \tilde{\Psi}_1 , \mathcal{N}[\eta]\right)+\left( \tilde{\Psi}_1 ,\sqrt{\frac{2D}{ N}} \frac{\partial  }{\partial \theta}\left[\sqrt{\frac{\delta(\omega)}{2\pi}+\eta(\theta,\omega,t)}\zeta(\theta,\omega,t)\right] \right). 
\end{eqnarray}
We have computed in Section~\ref{sec: sup 1} the first two terms, which become $\lambda_1A$ and $-c_3A|A|^2$ (in the leading order), respectively, where $c_3$ is given by Eq.~\eqref{eq: C3 App 1 SUP}. Defining the third term as noise $\mathbf{N}(t)$, we have
\begin{eqnarray}
    \mathbf{N}(t)&=&\frac{1}{2\pi}\sqrt{\frac{2D}{ N}}\int_{-\infty}^{\infty}d\omega\int_{0}^{2\pi}d\theta~ \tilde{\psi}^{*}_1(\omega)e^{-i\theta} \frac{\partial  }{\partial \theta}\left[\sqrt{\frac{\delta(\omega)}{2\pi}+\eta(\theta,\omega,t)}\zeta(\theta,\omega,t)\right]\nonumber\\
    &=& \frac{i}{2\pi}\sqrt{\frac{2D}{ N}}\int_{-\infty}^{\infty}d\omega\int_{0}^{2\pi}d\theta~ \tilde{\psi}^{*}_1(\omega)e^{-i\theta} \sqrt{\frac{\delta(\omega)}{2\pi}+\eta(\theta,\omega,t)}\zeta(\theta,\omega,t),
\end{eqnarray}
where we have performed integration by parts. Let us compute the properties of the noise. Clearly, we have $\langle \mathbf{N}(t)\rangle = 0$. Now, we have
\begin{eqnarray}
   && \langle \mathbf{N}(t)\mathbf{N}(t')\rangle \nonumber \\
    &=& -\frac{D}{2\pi^2 N}\int d\omega d\omega'd\theta d\theta'~ \tilde{\psi}^{*}_1(\omega)\tilde{\psi}^{*}_1(\omega')e^{-i(\theta+\theta')} \sqrt{\left[\frac{\delta(\omega)}{2\pi}+\eta(\theta,\omega,t)\right]\left[\frac{\delta(\omega')}{2\pi}+\eta(\theta',\omega',t')\right]}\langle\zeta(\theta,\omega,t')\zeta(\theta',\omega',t')\rangle \nonumber \\
    &=& -\frac{D}{2\pi^2 N}\int d\omega d\omega'd\theta d\theta'~ \tilde{\psi}^{*}_1(\omega)\tilde{\psi}^{*}_1(\omega')e^{-i(\theta+\theta')} \sqrt{\left[\frac{\delta(\omega)}{2\pi}+\eta(\theta,\omega,t)\right]\left[\frac{\delta(\omega')}{2\pi}+\eta(\theta',\omega',t')\right]}\delta(\theta-\theta')\delta(\omega-\omega')\delta(t-t')\nonumber\\
    &=& -\delta(t-t')\frac{D}{2\pi^2 N}\int_{-\infty}^{\infty}d\omega\int_{0}^{2\pi}d\theta~\left[\tilde{\psi}^{*}_1(\omega')\right]^2e^{-i2\theta}\left[\frac{\delta(\omega)}{2\pi}+\eta(\theta,\omega,t)\right]\\
    &=&  -\delta(t-t')\frac{D}{2\pi^2 N}\int_{-\infty}^{\infty}d\omega\left[\tilde{\psi}^{*}_1(\omega')\right]^2\int_{0}^{2\pi}d\theta~e^{-i2\theta}\eta(\theta,\omega,t),
\end{eqnarray}
where we have used $\int_{0}^{2\pi}d\theta e^{-i2\theta} = 0$. Using the Fourier expansion of $\eta$, we obtain
\begin{equation}
   \langle \mathbf{N}(t)\mathbf{N}(t')\rangle =  -\delta(t-t')\frac{D}{ \pi N}\int_{-\infty}^{\infty}d\omega\left[\tilde{\psi}^{*}_1(\omega')\right]^2W_2\left[A,A^{*}\right] = -\delta(t-t')\frac{DA^2}{\pi N}\int_{-\infty}^{\infty}d\omega\left[\tilde{\psi}^{*}_1(\omega')\right]^2w_{2,0}(\omega) + \mathcal{O}(A^2|A|^2).
\end{equation}
Now, putting the expressions of $\tilde{\psi}_1(\omega)$ and $w_{2,0}(\omega)$, we obtain
\begin{equation}
    \int_{-\infty}^{\infty}d\omega\left[\tilde{\psi}^{*}_1(\omega')\right]^2w_{2,0}(\omega) =\int_{-\infty}^{\infty}d\omega\left[\frac{K_1}{K_1+i2\omega}\right]^2 \frac{2\pi K_1 \delta(\omega)   }{\left(K_1 + 2D\right)}  = \frac{2\pi K_1    }{\left(K_1 + 2D\right)}.
\end{equation}
Hence, in the leading order, we have
\begin{equation}
   \langle \mathbf{N}(t)\mathbf{N}(t')\rangle =  -\frac{A^2}{N}\left[\frac{2 K_1D}{ K_1+2D}\right]\delta(t-t').
   \label{eq:correlation}
\end{equation}

In a similar way, we compute
\begin{eqnarray}
   && \langle \mathbf{N}(t)\mathbf{N}^{*}(t')\rangle \nonumber \\
    &=& \frac{D}{2\pi^2 N}\int d\omega d\omega'd\theta d\theta'~ \tilde{\psi}^{*}_1(\omega)\tilde{\psi}_1(\omega')e^{-i(\theta-\theta')} \sqrt{\left[\frac{\delta(\omega)}{2\pi}+\eta(\theta,\omega,t')\right]\left[\frac{\delta(\omega')}{2\pi}+\eta(\theta',\omega',t)\right]}\langle\zeta(\theta,\omega,t)\zeta(\theta',\omega',t')\rangle \nonumber \\
    &=& \frac{D}{2\pi^2 N}\int d\omega d\omega'd\theta d\theta'~ \tilde{\psi}^{*}_1(\omega)\tilde{\psi}_1(\omega')e^{-i(\theta-\theta')} \sqrt{\left[\frac{\delta(\omega)}{2\pi}+\eta(\theta,\omega,t)\right]\left[\frac{\delta(\omega')}{2\pi}+\eta(\theta',\omega',t')\right]}\delta(\theta-\theta')\delta(\omega-\omega')\delta(t-t')\nonumber\\
    &=& \delta(t-t')\frac{D}{2\pi^2 N}\int_{-\infty}^{\infty}d\omega\int_{0}^{2\pi}d\theta~\left|\tilde{\psi}_1(\omega')\right|^2\left[\frac{\delta(\omega)}{2\pi}+\eta(\theta,\omega,t)\right]\\
    &=&  \delta(t-t')\frac{D}{2\pi^2 N}\left[\int_{-\infty}^{\infty}d\omega\left|\tilde{\psi}_1(\omega')\right|^2\delta(\omega)\right]\left[\frac{1}{2\pi} \int_{0}^{2\pi}d\theta\right].
\end{eqnarray}
Hence, we have
\begin{equation}
    \langle \mathbf{N}(t)\mathbf{N}^{*}(t')\rangle = \left[\frac{D}{2\pi^2 N}\right]\delta(t-t'),
    \label{eq:correlationstar}
\end{equation}
The equation for $A$ is then
\begin{equation}
    dA= \left(\lambda_1A-c_3A|A|^2\right)dt+d\mathbf{N}(t),
\end{equation}
where the complex noise $\mathbf{N}$ is gaussian, centered, and determined by its correlations
\eqref{eq:correlation} and \eqref{eq:correlationstar}.

Now, we want to compute the evolution of $r = |A| = \sqrt{AA^{*}}$; this requires the use of Ito formula. For the sake of completeness, we give the whole computation. Taylor expansion up to the second order gives
\begin{equation}
    dr = \frac{\partial r}{\partial A} dA +\frac{\partial r}{\partial A^{*}} dA^{*} + \frac{1}{2} \left[\frac{\partial^2 r}{\partial A\partial A}dAdA+ 2\frac{\partial^2 r}{\partial A^{*}\partial A}dA^{*}dA+ \frac{\partial^2 r}{\partial A^{*}\partial A^{*}}dA^{*}dA^{*}\right].
\end{equation}
Now, we have
\begin{equation}
    \frac{\partial r}{\partial A} = \frac{A^{*}}{2r},~\frac{\partial r}{\partial A^{*}} = \frac{A}{2r},~\frac{\partial^2 r}{\partial A\partial A} = -\frac{r}{4A^2},~\frac{\partial^2 r}{\partial A^{*}\partial A}=\frac{\partial^2 r}{\partial A\partial A^{*}} =\frac{1}{4r},~\frac{\partial^2 r}{\partial A^{*}\partial A^{*}} =-\frac{r}{4\left(A^{*}\right)^2}.
\end{equation}
Considering terms up to $\mathcal{O}(dt)$, and using \eqref{eq:correlation} and \eqref{eq:correlationstar}, we obtain
\begin{eqnarray}
    dr &=& \frac{A^{*}}{2r}\left[\left(\lambda_1A-c_3A|A|^2\right)dt+d\mathbf{N}(t)\right]+\frac{A}{2r}\left[\left(\lambda_1A^{*}-c_3A^{*}|A|^2\right)dt+d\mathbf{N}^{*}(t)\right] \nonumber \\
          && +\frac{1}{2}\left[\frac{r}{4A^2}\frac{A^2}{N}\left[\frac{2 K_1D}{ K_1+2D}\right]+2\frac{1}{4r}\left[\frac{D}{2\pi^2 N}\right]+\frac{r}{4\left(A^{*}\right)^2}\frac{\left(A^{*}\right)^2}{N}\left[\frac{2 K_1D}{ K_1+2D}\right]\right]dt \\
          &=& \left\{\lambda_1 r-c_3r^3 + \frac{D}{8\pi^2 N r}\left[1+r^2\left(\frac{4\pi^2K_1}{K_1+2D}\right)\right]\right\}dt + \frac{1}{2r} \left[A^{*}d\mathbf{N}(t)+Ad\mathbf{N}^{*}(t)\right],
\end{eqnarray}
which gives
\begin{equation}
    \frac{dr}{dt} = \lambda_1 r-c_3r^3 + \frac{D}{8\pi^2 N r}\left[1+r^2\left(\frac{4\pi^2K_1}{K_1+2D}\right)\right] + \mathbf{N}_r(t),
\end{equation}
where $\mathbf{N}_r(t) = (2r)^{-1}\left[A^{*}\mathbf{N}(t)+A\mathbf{N}^{*}(t)\right]$. Since $A(t)$ and $\mathbf{N}(t)$ are independent, the correlation properties of the gaussian noise $N_r$ can be obtained from \eqref{eq:correlation} and \eqref{eq:correlationstar}; we obtain
\begin{eqnarray}
    \langle\mathbf{N}_r(t) \rangle = 0,~\langle\mathbf{N}_r(t)\mathbf{N}_r(t') \rangle  = \frac{D}{4\pi^2N}\left[1-r^2\left(\frac{4\pi^2K_1}{K_1+2D}\right)\right]\delta(t-t').
\end{eqnarray}
Hence, we may write
\begin{eqnarray}
    \mathbf{N}_r(t) = \sqrt{\frac{D}{4\pi^2N}\left[1-r^2\left(\frac{4\pi^2K_1}{K_1+2D}\right)\right]} \xi_r(t),
\end{eqnarray}
where we have $\langle\xi_r(t)\rangle = 0$ and $\langle\xi_r(t)\xi_r(t')\rangle = \delta(t-t')$. Hence, we get
\begin{equation}
    \frac{dr}{dt} = \lambda_1 r-c_3r^3 + \frac{D}{8\pi^2 N r}\left[1+r^2\left(\frac{4\pi^2K_1}{K_1+2D}\right)\right] +  \sqrt{\frac{D}{4\pi^2N}\left[1-r^2\left(\frac{4\pi^2K_1}{K_1+2D}\right)\right]} \xi_r(t).
\end{equation}
Substituting $r = R_1/(2\pi)$, we obtain
\begin{equation}
    \frac{dR_1}{dt} = \lambda_1 R_1-\frac{c_3}{4\pi^2}R_1^3 + \frac{D}{2 N R_1}\left[1+R_1^2\left(\frac{K_1}{K_1+2D}\right)\right] +  \sqrt{\frac{D}{N}\left[1-R_1^2\left(\frac{K_1}{K_1+2D}\right)\right]} \xi_r(t).
\end{equation}
The above is Eq.~(13) of the main text.

\end{document}